\documentclass[11pt]{article}

\usepackage[margin=1in]{geometry}
\usepackage[T1]{fontenc}
\usepackage[utf8]{inputenc}
\usepackage{lmodern}
\usepackage{microtype}

\usepackage{amsmath,amssymb,mathtools}
\usepackage{graphicx}
\usepackage{xcolor}
\usepackage[numbers,sort&compress]{natbib}
\usepackage[hidelinks]{hyperref}
\usepackage{authblk}

\title{Catastrophic disruption cascades driven by the nonlinearity of systemic risk}

\author[1,2]{Jan Fialkowski}
\author[3]{Shlomo Havlin}
\author[1,2,4]{Stefan Thurner\thanks{Corresponding author: \href{mailto:Stefan.Thurner@meduniwien.ac.at}{Stefan.Thurner@meduniwien.ac.at}}}

\affil[1]{Complexity Science Hub, Vienna, Austria}
\affil[2]{Institute of the Science of Complex Systems, CeDAS, Medical University Vienna, Vienna, Austria}
\affil[3]{Department of Physics, Bar-Ilan University, Ramat-Gan 52900, Israel}
\affil[4]{Santa Fe Institute, Santa Fe, USA}

\date{}

\begin{document}
\maketitle

\begin{abstract}
Whether the COVID-19 pandemic or the Iran war, recent events have highlighted the systemic fragility of supply chains. Due to highly specific and mutual buyer-supplier dependencies, even the failure of a single firm can cause system-wide economic disruptions in the form of cascading failures up and down the supply chain network. Only recently has it become possible to quantify the systemic impact of the failure of individual firms on the total supply chain. Here, we demonstrate that the systemic risk contributions of combinations of firm failures can be drastically larger than the sum of the damage caused by the firms individually. Using a unique data set that allows us to reconstruct the national supply chain network of Ecuador at the firm-level, we find that combined failures can produce systemic risk amplifications of up to a factor of 257. However, only a tiny fraction of 0.14\% of pairs exhibit a more than 4-fold amplification of systemic risk. 
We develop a simple method to identify firm combinations that lead to large systemic risk amplifications. The origin of these amplifications is a breakdown of the substitutability of defaulted suppliers. We discuss the implications of the existence of rare but strong systemic risk amplification for situations that simultaneously affect multiple firms, such as natural disasters and wars.
\end{abstract}


\section{Introduction}
In many networked systems, disruptions can propagate from a local point of failure through the network and potentially cause system-wide damage. This behavior can be observed in power outages in power grids \cite{Sch18}, in traffic jams \cite{Zha19}, or wildfires and landslides \cite{Tur04}. A typical feature of cascading disruptions is that the distribution of total damage is heavy-tailed, meaning that large-scale events occur way more often than what is expected from Gaussian statistics.

Systemic risk, the risk of system-wide cascading disruptions triggered by local events, has long eluded quantification, since the underlying network data have not been available. In the context of financial systemic risk, betweenness centrality was introduced as a simple systemic risk measure \cite{Bos04,Bos04a} that later were improved by measures that better capture the actual dynamics of cascading failures. For example, the so-called debt rank \cite{Bat12} models cascading financial losses by accounting for balance-sheet identities and accounting practices. If a bank experiences financial distress, the loans extended to it lose value. Under mark-to-market accounting, this devaluation results in a partial write-down of these loans on creditors' balance sheets \cite{Thu12}. This, in turn, can cause further financial distress among creditors, which may then propagate through the network of interbank loans.
The concept became particularly prominent after the global financial crisis of 2008 \cite{Gai10}, when the failure of a few interconnected financial institutions led to widespread financial distress and losses across the economy. Models such as DebtRank \cite{Bat12}, as well as related valuation-based models \cite{Bar15,Bar20}, have been used to study financial systemic risk in networks with multiple types of connections \cite{Pol15} and indirect contagion through overlapping portfolios \cite{Pol21,Pic21a,Con17}. Further work has shown that total systemic risk in a financial network can be minimized through simple network restructuring \cite{Thu13,Pic21,Die20}. Further, such models allow for generic insights into the stability of financial systems\cite{Bar17,Wie23}. For a review on this topic, we refer the interested reader to \cite{Bar21,Thu22}.

Supply chains are subject to a similar kind of fragility. When a firm fails and ceases production, its customers might face reduced input availability and may have to reduce their production levels as well. At the same time, its suppliers may face reduced demand and likewise reduce their production. These production losses can then propagate both upstream and downstream through the network of supplier-buyer relationships. Such cascading effects have been documented for idiosyncratic shocks \cite{Bar16a} and played a crucial role during the COVID-19 pandemic \cite{Pic21,Die24}, Hurricane Katrina \cite{Hal08}, and the 2011 earthquake in Japan \cite{Ino19,Car21}. Recent advances in the availability of firm-level data \cite{Bac23} have made it possible to model cascading disruptions in national economies at the level of individual firms. The economic systemic risk index of a firm, ESRI$_i$, estimates the fraction of total production lost as a consequence of that firm’s failure \cite{Die22}.

Here, we use a unique firm-level dataset covering the national economy of Ecuador. We reconstruct Ecuador’s complete national supply chain network from administrative VAT data for 2015. In addition, we extract smaller subnetworks centered on the crustacean and softdrink supply chain networks, given their importance for the national economy; see SI section~A.4. Details on the dataset and the extraction procedure are provided in Materials and Methods and in \cite{Zel26}. Using this supply chain network, we estimate how the disruption of a firm affects both its customers, through input shortages, and its suppliers, through demand losses. These customers and suppliers may then propagate the disruption further through the network \cite{Die22}. Quantifying the impact of a firm’s failure on its customers requires accounting for two mechanisms. First, firms’ production technologies are heterogeneous: some inputs are essential for production, whereas others are non-essential. Second, firms may be able to replace failing suppliers with other firms that produce similar products. However, such substitution is feasible only when the failing supplier is small compared to the relevant supplying industry, i.e., its market share is small. Details of the model are provided in Materials and Methods, see also \cite{Die22}.

Although ESRI is typically defined for the failure of individual firms, it can also be applied to larger disruption scenarios, such as the COVID-19 pandemic \cite{Die24}. One striking feature of firm-level economic systemic risk is the emergence of systemic risk plateaus in many supply chain networks. These plateaus consist of a relatively small set of firms whose failure disrupts a large fraction of the network. Both the full national supply chain network and the sofdrink supply chain network exhibit such plateaus. In the national network, a plateau of 64 firms reduces total production by more than 80\%; in the softdrink network, the corresponding plateau consists of 18 firms, see SI section~B.1
for details. In the context of DebtRank, losses from a large scenario—for example, the simultaneous failure of two banks—are at most equal to the sum of the losses caused by the corresponding individual failures. In this sense, DebtRank is sublinear by definition; see SI section~B.7. The additional complexity of supply chain networks and firm-level production processes raises the question of whether ESRI satisfies an analogous sublinearity property.

Here, we examine whether the total damage caused by the simultaneous failure of small sets of firms is similar to the sum of the damages caused by the failure of each firm individually. We denote the failure of a single firm by $\psi_i$ and the resulting fractional reduction in total network production by $\text{ESRI}(\psi_i)$. We denote a small set of firms by $\mathcal{S}_k$, and the simultaneous failure of all firms in such a set by $\sum_{i\in\mathcal{S}_k}\psi_i$. To quantify nonlinear effects, we define an amplification factor $\alpha$ as
\begin{equation}
\alpha = \frac{\text{ESRI}\left(\sum_{i\in\mathcal{S}_k}\psi_i\right)}{\sum_{i\in\mathcal{S}_k}\text{ESRI}\left(\psi_i\right)}\,.\label{eq:alpha}
\end{equation}
This amplification factor measures whether the systemic risk of a joint failure, $\text{ESRI}\left(\sum_{i\in\mathcal{S}_k}\psi_i\right)$, is larger or smaller than the sum of the corresponding individual systemic risks, $\sum_{i\in\mathcal{S}_k}\text{ESRI}\left(\psi_i\right)$. Values of $\alpha<1$ indicate sublinear effects. Our main interest, however, lies in sets with $\alpha>1$, which indicate nonlinear amplification of systemic risk.

Identifying sets of firms that exhibit nonlinear amplification is a combinatorial challenge. The full national supply chain network contains roughly $10^5$ firms, corresponding to about $10^{10}$ possible firm pairs, which makes exhaustive search infeasible. We address this challenge by first shocking large sets of firms and then extracting minimal subsets whose joint failure still gives rise to nonlinear amplification. This strategy is related to the random chemistry approach \cite{Kau94}, which has been successfully applied to power grids \cite{Epp12}. Because the softdrink subnetwork is much smaller, containing only 1075 firms, we can exhaustively examine all firm pairs in this network and use the results to validate the minimal-subset extraction strategy; see SI section~A.3.

\section{Results}

\begin{figure}
    \centering
    \includegraphics[width=0.5\linewidth]{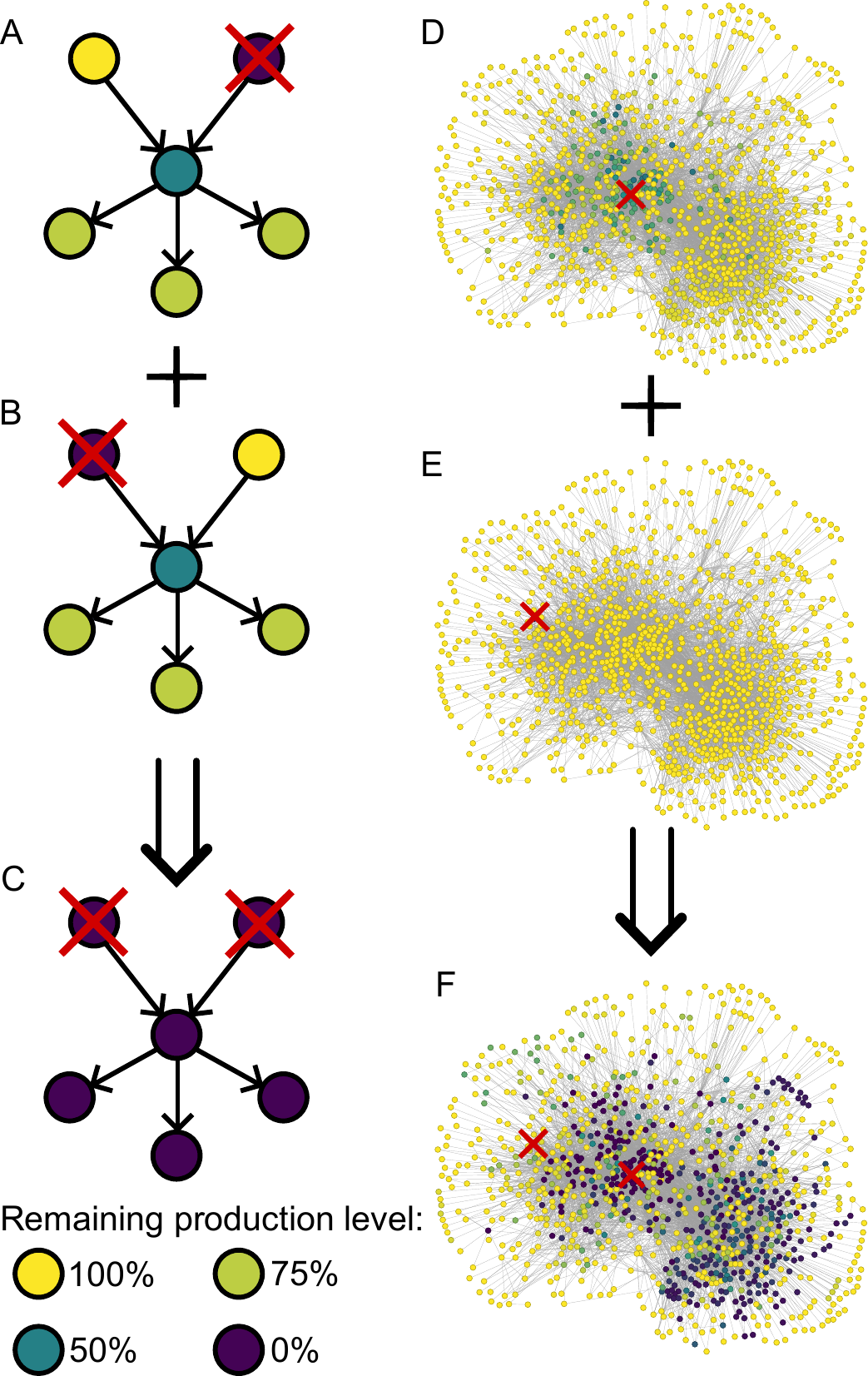}
    \caption{\textbf{Nonlinear amplification of systemic risk in a supply chain network.} The left column illustrates a schematic example of nonlinear amplification. In panel A, a single supplier firm is shocked, marked by the red cross. As a result, this firm completely stops production and is shown in dark purple. The shock then propagates through the network and reduces production at downstream firms. Yellow indicates no production loss for the second supplier, blue indicates a 50\% production reduction for the focal firm, and green indicates a 25\% production reduction for the three final nodes. After the initial shock has cascaded through the network, total damage amounts to 40\% of the total network production. Panel B shows the effect of the failure of the second supplier firm alone, which leads to a similar reduction in production among downstream firms. In panel C, both supplier firms fail simultaneously. The central firm must now stop production completely. Because the complete loss of inputs cannot be compensated for by the three final firms, they also cease production. Total damage therefore amounts to 100\% of network production, more than twice the damage caused by the failure of either supplier individually. The right column shows an example from the crustacean supply chain network. In panel D, a single firm fails, marked by the red cross. As a result, some firms reduce production, shown by the green dots. In panel E, a different firm fails, leading to a much smaller cascade. In panel F, both firms fail simultaneously, triggering a large cascade in which many firms must stop production completely, shown by the dark purple dots.
    }
    \label{fig:Figure1}
\end{figure}

In Fig.~\ref{fig:Figure1}~A--C, we show a schematic example of nonlinear amplification of systemic risk. In panels A and B, we shock a single supplier firm, marked by a red cross and shown as a dark purple node. As a result, the focal firm in the example partially reduces its production, which in turn causes small disruptions among the three downstream firms. The failure of a single supplier can be partially compensated for by the focal firm, as long as the other supplier remains available. The total damage caused by the failure of either supplier individually amounts to 40\% of the total network production. In panel C, both suppliers fail simultaneously. Because no alternative supplier remains available, the central firm can no longer compensate for the missing inputs, leading to network-wide losses that are larger than the sum of those caused by the individual failures of the suppliers.

In Fig.~\ref{fig:Figure1}~D--F, we show an empirical example from the Ecuadorian crustacean supply chain network. We shock two firms in the “Cargo Handling” sector, ISIC classification number H5224, whose individual failures cause only small disruptions, but whose simultaneous failure leads to a much larger cascade. This nonlinear amplification occurs because both firms become difficult to replace when their shared industry is disrupted. The firm in panel D has the largest market share in the sector, at 33\%, and its individual failure leads to a 6.5\% reduction in total network output. The firm in panel E is the fifth-largest firm, with a market share of 5\%, and has an individual ESRI of 0.2\%. The joint failure of these two firms leads to $\mathrm{ESRI}(\psi_i+\psi_j)=0.35$, which is more than five times as large as the sum of their individual impacts, $\mathrm{ESRI}(\psi_i)+\mathrm{ESRI}(\psi_j)=0.067$. This example also highlights that market share alone is not sufficient to predict nonlinear amplification. The simultaneous failure of the two firms with the largest market shares, 33.4\% and 4.7\%, respectively, leads to a loss of only 6.7\% of the total network output. This loss is not only smaller than in the previous scenario, but also smaller than the sum of the two firms’ individual ESRIs, $\mathrm{ESRI}(\psi_i)+\mathrm{ESRI}(\psi_j)=0.072$.

We now examine the extent and the causes of nonlinear amplification of systemic risk. We begin by studying nonlinear amplification in the crustacean and softdrink subnetworks, whose smaller size allows for an exhaustive analysis of all firm pairs. For the complete supply chain network of Ecuador, we examine a sample of two million firm pairs and apply our extraction procedure to efficiently identify small firm sets with large amplification factors, $\alpha$. Finally, we analyze the most strongly amplified firm sets and identify the structural motifs that give rise to large nonlinear amplification.

\subsection{Identification of nonlinearly interacting sets of firms}

\begin{figure}
    \centering
    \includegraphics[width=0.5\linewidth]{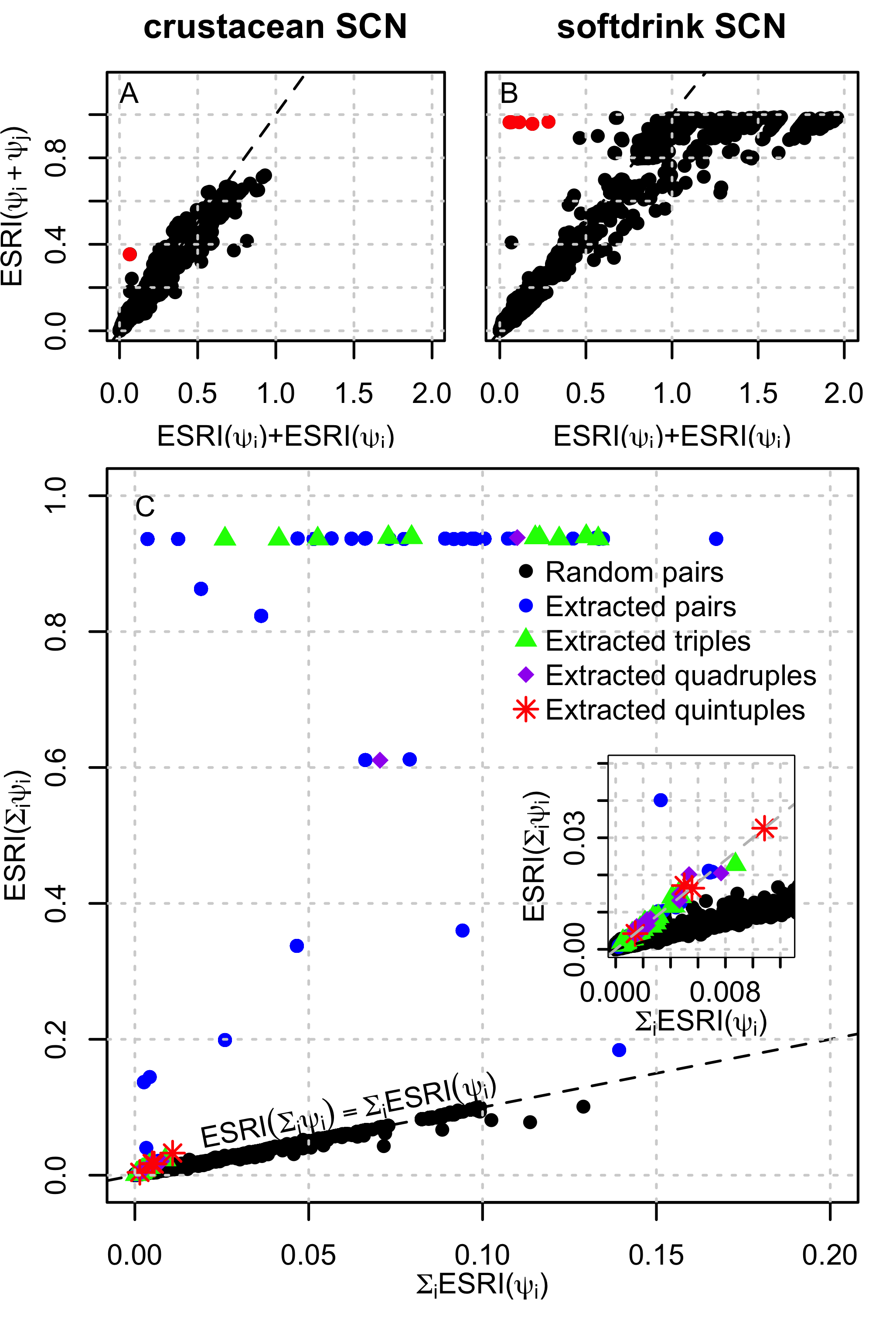}
    \caption{\textbf{Scatter plots of the ESRI for simultaneous firm failures versus the sum of individual firms' ESRIs.}
    Panels A and B show the ESRI of simultaneous failure, $\text{ESRI}(\psi_i+\psi_j)$, against the sum of the individual ESRIs, $\text{ESRI}(\psi_i)+\text{ESRI}(\psi_j)$, for all possible pairs in the crustacean (A) and softdrink supply chain network (B) respectively. The dashed diagonal indicates equality between the ESRI of the simultaneous failure and the sum of the individual ESRIs. In the crustacean supply chain network, most pairs are clustered around the diagonal, indicating that the damage caused by simultaneous failure is comparable to the sum of the damages caused by the individual failures. Only a few pairs exhibit clearly sublinear or superlinear behavior. The red dot highlights the example shown in Fig.~\ref{fig:Figure1}. In the softdrink supply chain network, panel B, many pairs exhibit sublinear behavior, i.e.,  they lie below the diagonal.
    Conversely, a small number of pairs lie above the diagonal: these pairs have low individual ESRIs but cause large systemic damage when they fail simultaneously. In particular, red dots mark six firm pairs with an amplification factor, $\alpha$, larger than three, whose simultaneous failure affects almost the entire supply chain network. Panel C shows the corresponding results for the complete national supply chain network of Ecuador. Black dots represent a random sample of 2,000,000 firm pairs. Colored symbols indicate small firm sets identified by our extraction procedure. Our method identifies pairs (blue dots), triples (green triangles), and quadruples (purple diamonds) whose simultaneous failure can affect the failure of almost the entire network, $\text{ESRI}(\sum_i \psi_i)\approx 1$, even though any individual firm has an ESRI below 10\%. The inset shows a blowup of the region near the origin. There, we identify quintuples (red stars) with an amplification factor, $\alpha$, close to 3. This corresponds to our detection threshold $\theta_1=3$, indicated by the gray dashed line. For our extraction procedure, we use 100,000 sets of $N=500$ and $N=100$ initial firms, and choose as thresholds $\theta_1=3$ and $\theta_2=0.9$; see Materials and Methods.
    }
    \label{fig:Figure2}
\end{figure}

Figure~\ref{fig:Figure2}~A shows, for all firm pairs in the crustacean subnetwork, the systemic risk of their simultaneous failure, $\text{ESRI}(\psi_i+\psi_j)$, versus the sum of the individual risks, $\text{ESRI}(\psi_i)+\text{ESRI}(\psi_j)$. If systemic risk would be strictly additive, all pairs would lie on the dashed line. In the crustacean supply chain network, most pairs lie close to this line, indicating that their joint impact is nearly additive. However, there are notable outliers. Pairs below the diagonal are jointly less destructive than the sum of their individual failures, for example, when the two cascades affect the same firms through different essential inputs. Under the Leontief production function, a 50\% shock to two essential inputs has the same effect on output as a 50\% shock to only one of them. Overall, nonlinear amplification in the crustacean supply chain network is limited. One of the strongest amplified outliers, marked in red, is the pair presented in Fig.~\ref{fig:Figure1}~D--F.

Figure~\ref{fig:Figure2}~B shows $\text{ESRI}(\psi_i+\psi_j)$ versus $\text{ESRI}(\psi_i)+\text{ESRI}(\psi_j)$ for all firm pairs in the softdrink subnetwork. Two patterns are obvious. Many pairs lie below the diagonal. This reflects the presence of firms whose individual failure already disrupts substantial parts of the network, so that an additional firm failure often increases the total damage marginally. Second, six pairs appear in the upper left corner of the plot and are marked in red. These pairs reduce the total output of the softdrink supply chain network by more than 90\%, even though the sum of their individual ESRIs is less than 30\%. One of them exhibits an amplification factor, $\alpha$, greater than 17. Thus, while strong nonlinear amplification is rare, it can have substantial consequences for the network. In contrast to the crustacean subnetwork, the softdrink subnetwork exhibits much stronger nonlinear amplification of systemic risk.

Since both the softdrink subnetwork and the full Ecuadorian supply chain network contain a plateau of highly risky single firms, the full network is a natural setting in which to search for strongly amplified firm sets. Although such sets are rare, their potential to cause disproportionate network-wide disruptions makes their systematic identification important. Because the number of possible firm sets is combinatorially large, exhaustive search is infeasible even when restricted to pairs. We therefore complement a random sample of two million firm pairs with sets identified by our extraction procedure, described in Materials and Methods. In this analysis, we exclude 95 firms with $\text{ESRI}(\psi_i)>0.1$, so that the search focuses on amplification arising from sets of firms that are not already highly systemically risky on their own.

Figure~\ref{fig:Figure2}~C compares two search strategies in the full Ecuadorian supply chain network: a random sample of two million firm pairs (black dots) and the sets identified with our extraction procedure (colored symbols). The inset shows a blow-up of the region near the origin. Random sampling does not identify any pairs whose simultaneous failure causes a strong systemic disruption of the network. By contrast, our extraction procedure uncovers sets of two, three, four, and five firms whose joint failure significantly impacts the supply chain network.
In total, it identifies 86 pairs (blue dots), 53 triples (green triangles), 17 quadruples (purple diamonds), and 4 quintuples (red stars).
Some of the identified sets are close to the origin and therefore have only a small aggregate impact on the economy, as shown in the inset. Note that it is possible for sets with a small $\text{ESRI}$ to have large amplification factors, $\alpha$. Several sets trigger cascades that disrupt nearly the entire network, with $\text{ESRI}(\sum_{i\in\mathcal{S}_k}\psi_i)\approx 1$. In particular, we find 28 pairs, 10 triples, and one quadruple whose simultaneous failure yields $\text{ESRI}(\sum_{i\in\mathcal{S}_k}\psi_i)>0.9$, even though the median sum of their individual ESRIs, $\sum_{i\in\mathcal{S}_k}\text{ESRI}(\psi_i)$, is only 0.094. The strongest nonlinear amplification is produced by a pair of firms whose individual ESRIs sum to only 0.0036, but whose simultaneous failure causes $\text{ESRI}(\psi_i+\psi_j)=0.94$, corresponding to an amplification factor of 257.

Strongly amplified firm sets are not unique to the 2015 Ecuadorian supply chain network. Applying the same extraction procedure to the 2010 network yields similar results; see SI section~B.3. Three of the highly amplified pairs identified in the 2015 network also appear in the 2010 network.

\begin{figure}[h]
    \centering
    \includegraphics[width=0.5\linewidth]{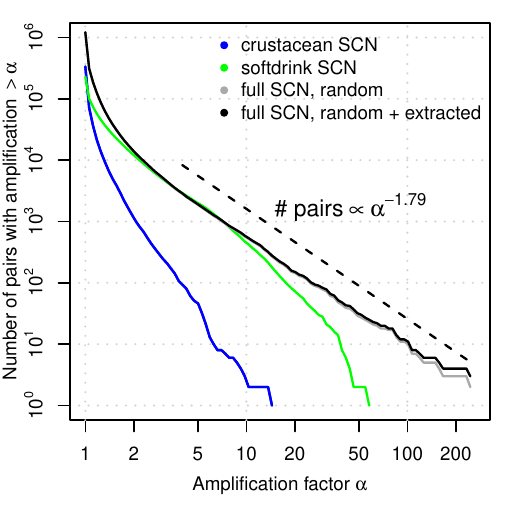}
    \caption{\textbf{Number of pairs with amplification factor larger than $\alpha$.} We show the number of firm pairs whose amplification factor, $\alpha$, exceeds a given threshold. The blue curve represents all firm pairs in the crustacean subnetwork, where amplification factors decay rapidly and reach a maximum value of $\alpha=15$. The green curve represents all firm pairs in the softdrink network. Here, amplification factors decay more slowly and reach values of up to $\alpha=59$. The light-grey curve represents two million randomly sampled firm pairs from the full supply chain network. The bulk of this distribution behaves similarly to that of the softdrink subnetwork, but its tail is substantially heavier: ten pairs exhibit amplification factors, $\alpha>100$, with a maximum value of $\alpha=257$. The dashed line shows a power law fit, following \cite{Cla09}, with an exponent of 1.79 for amplification factors larger than 4. The black curve additionally includes the firm pairs identified by our extraction procedure, which slightly increases the weight of the tail. Note that large amplification factors can occur even when the total impact of a firm pair is small.
    }
    \label{fig:Figure3}
\end{figure}

In Fig.~\ref{fig:Figure3}, we compare the number of pairs with a given amplification factor larger than $\alpha$ across the different networks. The crustacean network (blue) shows a rapid decay with a maximum amplification of $\alpha=15$. By contrast, the softdrink network (green) decays more slowly and reaches values up to $\alpha=59$, consistent with the stronger nonlinear interactions observed in Fig.~\ref{fig:Figure2}~B. Random sampling of firm pairs in the full Ecuadorian supply chain network (gray) shows that most pairs have amplification factors around 1, indicating additivity. Among the two million sampled pairs, 2836 have $\alpha>4$, which is $0.14\%$ of the randomly sampled pairs. A power-law fit following \cite{Cla09} yields a survival-function exponent of $-1.79$, indicating a broad distribution of amplification factors. Including the pairs identified by the extraction method extends the upper tail of the distribution. Most superlinear pairs found by random sampling have a very small systemic impact. Restricting attention to economically relevant pairs with $\text{ESRI}>0.01$ shows that only the extraction method identifies pairs that are both economically relevant and strongly amplified, reaching values up to $\alpha=257$; see SI section~B.5. Due to the rarity of strong non-linear amplification, the effect on the average systemic risk of pairwise firm failures is an increase of only 0.71\%, see SI section~B.4.

\subsection{Mechanisms for nonlinear amplification}

\begin{figure}[h!]
    \centering
    \includegraphics[width=0.5\linewidth]{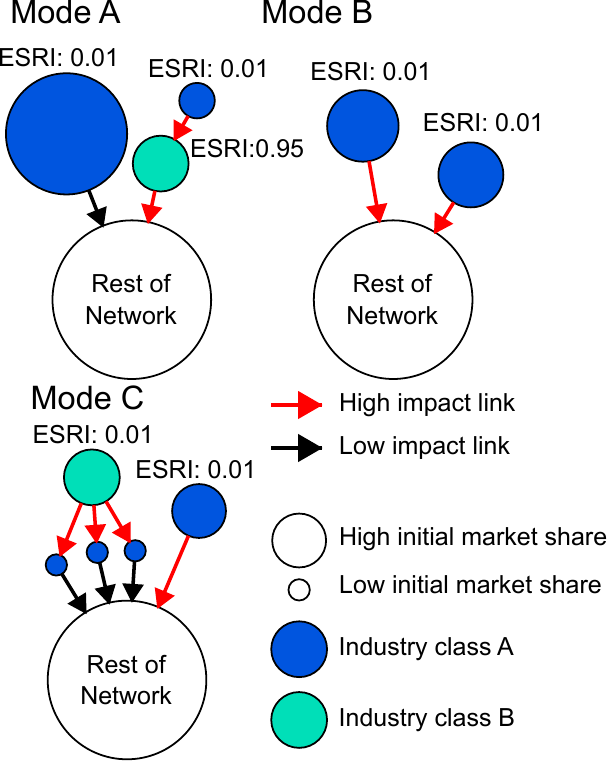}
    \caption{
    \textbf{Schematic network configurations leading to strong nonlinear amplification of systemic risk.} We show three representative network modes observed among firm pairs with large systemic risk. Color denotes the industry classification of a node, and its size denotes the initial market share. Red links denote buyer-supplier relationships that can completely disable the buyer if the supplier fails and cannot be replaced, i.e., high impact links with $\Lambda_{ij}=1$. Black links are relationships with limited impact on the buying node, i.e., $\Lambda_{ij}<1$. The mode in panel A consists of two nodes in the same industry, one with a large initial market share, but only a limited impact on its customers, and a small firm that serves as a high impact supplier to a "plateau-firm". In mode B, the two nodes share an industry, with a large combined market share and high impact links to many customers. Panel C shows a mode with a pair of firms in different industries. Here, one of the firms affects a large part of the other firms' industry, increasing its effective market share, $\sigma_i(t)$, see Eq.~\ref{eq:Sigma}, thus rendering it irreplaceable.}
    \label{fig:Motiveschemas}
\end{figure}

To understand the mechanisms behind strong nonlinear amplification, we examine the most destructive firm pairs in the full supply chain network and identify recurring structural motifs. We focus on representative examples that recur among the most harmful pairs. These motifs help clarify the structural conditions under which strong amplification is likely to arise.

Figure~\ref{fig:Motiveschemas} illustrates three representative structural modes that recur among the most disruptive pairs. The examples are drawn from analyzing the 32 pairs with $\text{ESRI}\left(\psi_i+\psi_j\right)>0.5$. Node size represents the initial market share, node color indicates industry, and high impact links, those with $\Lambda_{ij}=1$, are shown in red. Through these links, a failing supplier can completely disrupt the production of its customer if no alternative supplier is available. The exact weight of each link is not relevant to understand these examples and thus not shown. The first mode, shown in Fig.~\ref{fig:Motiveschemas}~A, consists of a small firm that acts as a high impact supplier to a high-risk firm and a second firm with a large market share in the same industry as the small supplier. If the small supplier fails on its own, the resulting shock to the "plateau-firm" is attenuated because the supplier’s initial market share is low; see Eq.~\ref{eq:Sigma}. However, if the small supplier fails simultaneously with the large firm in the same industry, the shock can no longer be attenuated because replacement becomes more difficult. The "plateau-firm" is then strongly disrupted and initiates a cascade that affects large portions of the network. The large-market-share firm need not be directly connected to the "plateau-firm" or to the smaller supplier beyond belonging to the same industry. This mechanism is exemplified by the pair with the strongest amplification factor, $\alpha=257$, see SI section~B.6
for details.

In the second mode, shown in Fig.~\ref{fig:Motiveschemas}~B, the two initially shocked firms together account for a large share of their industry and also serve as high impact suppliers to many downstream firms. The failure of each one of them can still be compensated, but the simultaneous failure of both disrupts their industry strong enough that the supplier replaceability mechanism can no longer attenuate the shock. As a result, both shocks propagate downstream and affect large parts of the network. One example is the one shown in Fig.~\ref{fig:Figure1}~D--F. Another example of a pair found in the complete supply chain network of Ecuador is presented in SI section~B.6.

Note that the distinction between these two modes, A and B, is not always obvious. The structural motifs discussed above should be understood as representative patterns rather than mutually exclusive categories. For example, the smaller firm in the first mode may also have a number of high impact links to the wider network in addition to the one to the "plateau-firm". Conversely, in the second mode, one of the two firms may also act as a high impact supplier to a "plateau-firm".

The two initially shocked firms do not have to belong to the same industry to exhibit nonlinear amplification. We identify a third mode, shown in Fig.~\ref{fig:Motiveschemas}~C, where the two firms belong to different industries, but their cascades interact because both affect the same downstream industry. This interaction is observed in 17 of the 32 examined cases. The broad disruption of a shared downstream industry can disable the supplier replaceability mechanism and thereby allow the cascade to spread further through the network. An example of this mechanism is presented in SI section~B.6.

The supplier replaceability mechanism acts as a compensatory mechanism whose breakdown can strongly amplify losses in the network. This distinguishes our setting from other systemic risk frameworks that do not include such a mechanism. For example, DebtRank lacks an analogous compensatory channel and thus does not exhibit nonlinear amplification, and the impact of simultaneous bank failures combines sublinearly; see SI section~B.7.

\section{Discussion}
We show that the simultaneous failure of few firms with individually low systemic risk can give rise to output losses that are up to 257 times larger than the sum of their individual systemic risks. This massive nonlinear amplification arises when the core shock compensation mechanism, the ability to replace failing suppliers, fails. The disruption caused by a single failing supplier can usually be absorbed by finding an alternative source for the missing input, but the simultaneous failure of multiple firms can disrupt the supplying industry to the point that no alternative suppliers remain available. Supply chain networks can therefore be robust to the disruption of individual firms, yet fragile to the disruption of specific sets of a few firms.

While massive nonlinear amplification is rare, identifying sources of potentially catastrophic risks is paramount to securing the economy of a country against disruptions.
While the presence of nonlinear amplification increases the average systemic risk of pairwise firm failure by only 0.71\%, see SI section~B.4, it is a source of tail risk that can be exploited by targeted attacks or realized by large scale shocks. Nonlinear amplification becomes especially important when shocks do not affect firms at random, but instead disrupt the ability of supply chain networks to attenuate disruptions.

Empirically, we find small, rare sets of firms whose joint failure exhibits strong nonlinear amplification of systemic risk. In the reconstructed, national supply chain network, we identify small sets of firms whose simultaneous failure can reduce total network output by up to 90\%. The analysis of smaller subnetworks further shows that nonlinear amplification is stronger in networks that contain a systemic risk plateau, that is, a set of firms whose individual failures already cause large output losses. The simultaneous failure of multiple firms can affect these "plateau-firms," which we identify as a recurring motif that gives rise to massive nonlinear amplification of systemic risk. More generally, we identify three different structural motifs associated with nonlinear amplification due to pairwise firm failures. The two initially failing firms do not need to belong to the same industry. Instead, strong amplification often arises when their failure cascades jointly disrupt a shared downstream industry strongly enough that supplier replaceability breaks down. In such cases, the failure of one firm can substantially impair the industry in which the second firm operates, so that their simultaneous failure produces a much larger cascade than either failure would produce alone. These motifs can serve as a starting point for targeted search algorithms that identify potentially dangerous firm pairs.

We present a sampling procedure that efficiently identifies small firm sets whose joint failure exhibits strong nonlinear amplification of systemic risk. By initially shocking large sets of firms, the method probes many candidate subsets at once and extracts smaller sets that exhibit strong amplification of systemic risk. This approach is likely to be useful in other settings where nonlinear amplification is rare but can have disproportionately large effects compared with individual failures. A related method has previously been applied to power grids \cite{Epp12}, while other infrastructure networks, such as the internet, may provide further areas of application.

While ESRI is transparent and simple enough to be fully parameterized with available data, it makes some simplifying assumptions that may affect the quantitative results. It uses a stylized, market-share-based heuristic to estimate how easily a firm can replace a failing supplier, but it does not explicitly model rewiring to new suppliers. It also omits other mechanisms that can shape the propagation of shocks through the supply chain network. One of the most important missing mechanisms is price adjustment. Systemically important firms may be able to outbid others for scarce inputs, allowing them to continue operating and thereby reducing the overall impact of a cascading shock. The model also disregards the time dynamics of both the supply chain network and the shock propagation process. The available data are aggregated on a yearly level and do not contain information on inventories, production times, or short-run supplier adjustments, even though supply chain networks are known to be highly dynamic \cite{Rei26}. As a result, we cannot estimate how quickly disruptions propagate or how rapidly firms can adjust by finding alternative suppliers. The large-scale cascades we identify may therefore unfold only slowly, possibly over months. In this case, our results should be interpreted as an upper bound on the potential damage of a shock scenario. Studying the temporal profile of such cascades would require a dynamic systemic risk model that incorporates inventories, supplier rewiring, and time-varying network structure. Such model adjustments would likely change the quantitative results. However, as long as supply chain adjustments to a shock can fail under sufficient strain, nonlinear amplification should persist.

A second limitation concerns the available data. We lack information on international trade and final demand. These channels are important because imports can alleviate supply-side disruptions within the national supply chain network, while export exposure and final demand influence which firms and sectors are most relevant for aggregate output losses or as potential suppliers. Although we use firm-level data, we do not have access to product-level information on what firms actually produce. Such information would allow a more realistic treatment of input requirements and supplier replaceability. It may become available through invoice data \cite{Sil26,Di22a}. Alternatively, recent LLM-based approaches have begun to reconstruct product-level production networks \cite{Fet24}.

Finally, the presented set-extraction procedure is stochastic and cannot guarantee that we identify all firm sets that exhibit nonlinear amplification of systemic risk. For pairs, any given pair has a 96.8\% chance of being included in at least one of the larger sampled sets. For sets of three and more firms, however, the coverage drops drastically. Nevertheless, the method is sufficient to identify strongly amplifying sets and recover the structural motifs that give rise to nonlinear amplification.

Monitoring supply chain networks for systemic risk must account for nonlinear interactions, especially in networks that are already vulnerable to single-firm failures. Nonlinear amplification can create hidden vulnerabilities: firms that appear to have little systemic relevance in isolation may become systemically important when they fail together. Such hidden vulnerabilities may be particularly relevant for targeted attacks and correlated large-scale disruptions, such as natural disasters, geopolitical shocks, or trade restrictions, where failures are unlikely to occur independently at random. The data and methods presented in this paper provide a way to systematically identify these high-risk combinations. Thereby, it can enable more advanced systemic risk monitoring, as well as the design of prevention and mitigation strategies.

\section{Materials and Methods}

\subsection{Supply chain network data}
To reconstruct the national supply chain network of Ecuador, we use value-added tax data collected in 2015. The data provide the aggregated transaction value between all registered firms in the country. We encode these transactions in a weighted adjacency matrix, $W$, where each entry $W_{ij}$ denotes the total value of goods sold by firm $i$ to firm $j$. After removing all entities belonging to the “personas naturales” tax group, the full network contains 86,385 firms and 3,373,861 buyer–supplier links. From firms’ financial statements, we also extract the 4-digit ISIC classification, $p_i$, of each firm $i$.

From this national network, we extract smaller subnetworks centered on specific industries. We first identify a set of 109 seed nodes belonging to industry class C1020, “Processing and Preserving of Fish, Crustaceans, and Molluscs.” We then examine all direct suppliers of these seed nodes. We rank their 3-digit industry classifications by their overrepresentation relative to the full network and retain the sixteen most prominent industries that contain at least five firms. Using the same procedure, we select customers belonging to the eight most overrepresented industries. The crustacean subnetwork then consists of all selected nodes and the links between them. This procedure yields a network with 1,075 nodes and 5,969 edges.

We apply the same procedure to firms in category C1107, “Manufacture of Soft Drinks; Production of Mineral Waters and Other Bottled Waters,” to extract a softdrink supply chain subnetwork with 890 nodes and 6,040 edges. We focus on these two subnetworks because of their importance to the Ecuadorian economy: crustaceans and seafood account for a large share of Ecuador’s exports, while the soft-drink industry contributes around 35\% of value added in the manufacturing sector; see also SI section~A.4
and \cite{Zel26}.

\subsection{Supply chain shock propagation}
We use the ESRI model \cite{Die22} as the shock-propagation mechanism. We provide a brief overview of the model, focusing on supply-side shock propagation, and refer the reader to SI section~A.1
for full details. From the tax data, we reconstruct the weighted adjacency matrix and industry classification of each firm. For each product, we use the expert survey, conducted in \cite{Pic21}, to distinguish between essential and non-essential inputs for production. This information allows us to define the downstream shock propagation matrix, $\Lambda_{ij}^d$, which determines how strongly the failure of supplier $i$ affects its immediate customer $j$. The entries of $\Lambda_{ij}^d$ lie in $[0,1]$. A value of $\Lambda_{ij}^d=1$ means that, for example, a 50\% reduction in the production of supplier $i$ can lead to a 50\% reduction in the production of customer $j$. We call an entry with $\Lambda_{ij}^d=1$ a high impact link, because the failure of supplier $i$ can, in principle, completely stop production at firm $j$.

An important feature of the ESRI model is that suppliers can be replaced. Small suppliers are generally easier to replace than large suppliers, but replacement becomes more difficult when a shock affects a large fraction of the supplier's industry. To capture this mechanism, we define a supplier-replaceability factor that attenuates the propagation of shocks:
\begin{equation}
    \sigma_i\left(t\right)=\min\left[\frac{s_i}{\sum_j s_j\delta_{p_i p_j}\left(1-x_j^d(t)\right)},1\right]\,.
    \label{eq:Sigma}
\end{equation}
Here, $s_i$ denotes the total value of goods sold by firm $i$, i.e., its out-strength, $s_i=\sum_j W_{ij}$. The variable $x_i^d(t)$ denotes the fraction of production lost by firm $i$ due to input shortages at time $t$. At time $t=0$, before any firm has been affected by a shock, $\sigma_i(0)$ is equal to the market share of firm $i$ within its industry. We therefore refer to $\sigma_i(t)$ as the effective market share of firm $i$ at time $t$. The time dependence of $\sigma_i(t)$ is crucial: even a small firm can become difficult to replace if the rest of its industry is strongly affected by a shock.

We can now compute how an initial shock to firms' production capacities, $\psi$, propagates through the network and causes supply-side constraints for other firms:
\begin{multline}
        x_j^d(t+1) = \max\Biggl[\max_{p_l\in \mathcal{I}^\text{es}_j}\left[\sum_i \Lambda^d_{ij}x_i^d\left(t\right)\sigma_i\left(t\right)\delta_{p_i p_l}\right]\,,\\
    \sum_{p_l\in\mathcal{I}^\text{ne}_j}\sum_i \Lambda^d_{ij}x_i^d(t)\sigma_i(t)\delta_{p_i p_l},\psi_j\Biggr]\,.
    \label{eq:downprop_main}
\end{multline}
The entry $\psi_j$ of the initial shock vector, $\psi$, denotes the direct shock to the production capacity of firm $j$. A value of $\psi_j=1$ means that firm $j$ completely ceases production as a consequence of the initial shock. The first term captures constraints arising from essential inputs, where the most severe input shortage determines the production loss. The second term captures constraints arising from non-essential inputs, whose effects accumulate additively. The third term ensures that the direct initial shock to firm $j$ is retained throughout the propagation process. Note the nonlinearity introduced by the effective market share, $\sigma_i(t)$. The propagation of shocks to suppliers due to reduced demand follows a similar equation; see SI section~A.1
for details.

The propagation equation, Eq.~\ref{eq:downprop_main}, is iterated until the change in production capacity of each firm falls below $\epsilon=0.01$. The total effect of a given shock, $\psi$, on the network is summarized by the sum of the production losses of all firms, relative to the total production of all firms before the shock. This quantity is referred to as $\text{ESRI}(\psi)$. The shock vector, $\psi$, can encode arbitrary firm-level disruption scenarios. A shock vector in which only the $j$'th entry equals one, while all other entries are zero, represents the complete failure of firm $j$. Multiple simultaneous firm failures can be represented by adding the corresponding single-firm shock vectors. The resulting vector, therefore, encodes a scenario in which all selected firms are shocked simultaneously.

\subsection{Identifying sets with strong risk amplification}
We are interested in identifying small sets of firms, such as pairs and triplets, in which each individual firm has low systemic risk, as measured by ESRI, but whose simultaneous failure leads to large system-wide damage. We refer to this effect as nonlinear amplification. To identify such sets, we use an approach related to random chemistry. This method was first introduced in \cite{Kau94} and has been used, for example, to identify higher-order contingencies in power grids \cite{Epp12}. The approach begins by identifying large sets, in our case hundreds of firms, that exhibit the desired property: low individual ESRIs but high ESRI under simultaneous failure. Once such a set is found, firms are systematically removed until a minimal subset remains that still gives rise to strong nonlinear amplification.

First, we identify large firm sets that exhibit nonlinear amplification. We draw $N$ firms uniformly at random to form a set $\mathcal{S}_k$. We use $N=500$, and $N=100$ and generate 100,000 different sets for each size. From 86,385 firms in the full supply chain network, we exclude 95 firms with an individual systemic risk, $\text{ESRI}(\psi_i)>0.1$, leaving us with 86,290 candidate firms. For each set, $\mathcal{S}_k$, we calculate the ESRI of the simultaneous failure of all firms in the set and compare it with the sum of the individual firm ESRIs. We classify a set, $\mathcal{S}_k$, as successful if the ESRI of the simultaneous failure exceeds the sum of individual ESRIs by at least a factor of $\theta_1$,
\begin{equation}
    \text{ESRI}\left(\sum_{i\in \mathcal{S}_k}\psi_i\right)\geq \theta_1\sum_{i\in \mathcal{S}_k}\text{ESRI}\left(\psi_i\right)\,.
    \label{eq:StepOne}
\end{equation}
For our main results, we set $\theta_1=3$.
The candidate firms have an average ESRI of $1.5\times10^{-4}$. Thus, for sets of 500 firms and $\theta_1=3$, the threshold is, on average, 0.23. We perform this calculation for all 200,000 sampled sets and find 130 successful sets of 500 firms, and 105 successful sets of 100 firms.

From each successful set, we then extract the small subset of firms responsible for nonlinear amplification. To do so, we examine each of the $N$ possible subsets obtained by removing a single firm, $\mathcal{S}_k\setminus j$. For each subset, we calculate its ESRI and compare it with the ESRI of the full set of $N$ firms. We identify firms, $j$, whose removal leads to a substantial reduction in systemic risk, defined by
\begin{equation}
    \text{ESRI}\left(\sum_{i\in \mathcal{S}_k\setminus\{j\}}\psi_i\right)<\theta_2 \text{ESRI}\left(\sum_{i\in \mathcal{S}_k}\psi_i\right)\,.
\end{equation}
Here, $\theta_2$ controls how much the ESRI of the subset must decrease relative to the full set. We use a conservative threshold of $\theta_2 = 0.9$, meaning that removing a single firm from the large set must reduce ESRI by at least 10\%.

We examine the sensitivity of our results to each parameter in SI section~A.2. The number of firms in the initial sampled sets controls how effectively the search space is explored. Across 100,000 independently sampled sets of 500 Firms, the probability that any given firm pair, out of more than 3.6 billion possible pairs, is included in at least one sampled set is approximately 96.8\%.
However, using very large initial sets can obscure firm pairs with a small absolute ESRI but large amplification factors. Similarly, if $\theta_1$ is chosen too large, the procedure may miss sets with small ESRI, even if they exhibit strong amplification. By contrast, sets with large ESRI are identified robustly across a broad range of $\theta_1$ values. The parameter $\theta_2$ has little effect on the results. Finally, we validate the method on the softdrink supply chain network, where exhaustive enumeration is possible, and confirm that it identifies all relevant firm pairs; see SI section~A.3.

\section*{Data and code availability}
Due to legal restrictions, we are unable to share the VAT data for Ecuador. The code is available at \url{https://github.com/janfialkowski/NonlinearCascades_Repro}.

\section*{Author contributions}
All authors contributed to the design of the research. J.F. performed the research. All authors wrote the manuscript.

\section*{Competing interests}
The authors declare no competing interests.

\section*{Acknowledgments}
We thank Pablo Astudillo-Estevez for providing access to the supply chain data of Ecuador. The work of J.F. was funded by the Austrian Science Fund (FWF) under grant no. 10.55776/I5985 and 10.55776/P34994. The work of S.T. was funded by the Austrian Science Fund (FWF) under grant no. 10.55776/EFP5.

\newpage
\appendix

\section{Supplementary Materials and Methods}
\subsection{Full description of the ESRI model}\label{sec:ESRI}
We use the ESRI model \cite{Die22} as the shock-propagation mechanism. The weighted adjacency matrix $W$ represents the monetary value of goods sold between firms, with entry $W_{ij}$ denoting the value of goods sold by firm $i$ to firm $j$. We proxy the product of firm $i$, denoted by $p_i$, by its 4-digit ISIC classification. For each product, we use the expert survey conducted in \cite{Pic21} to distinguish between essential inputs, $\mathcal{I}^{\text{ess}}_j$, and non-essential inputs, $\mathcal{I}^{\text{ne}}_j$, required by firm $j$. We then define the downstream shock propagation matrix, $\Lambda_{ij}^d$, which determines how strongly the failure of supplier $i$ affects its immediate customer $j$, as follows:
\begin{align}
    \Lambda_{ij}^d&=\begin{cases}
        0 \qquad \text{ if } W_{ij}=0\,,\\
        \Lambda_{ij}^{d1} \quad \text{ if } p_i\in\mathcal{I}_j^\text{es}\,,\\
        \Lambda_{ij}^{d2} \quad \text{ if } p_i\in\mathcal{I}_j^\text{ne}\,,\\
    \end{cases}\\
    \Lambda_{ij}^{d1} &= \frac{W_{ij}}{\sum_l \delta_{p_l p_i}W_{lj}}\label{eq:LES}\,,\\
    \Lambda_{ij}^{d2} &= \frac{W_{ij}}{\sum_l W_{lj}}\label{eq:LNE}\,.
\end{align}
Here $W_{ij}$ is the weighted adjacency matrix of the supply chain network, $p_i$ the ISIC classification of a firm with the essential $\mathcal{I}^\text{ess}_i$ and non-essential inputs $\mathcal{I}^\text{ne}_i$.
Note the different denominators for essential and non-essential inputs. For essential inputs, the denominator contains only suppliers that provide the same product as supplier $i$. For non-essential inputs, by contrast, the denominator contains all inputs purchased by firm $j$. The entries of $\Lambda_{ij}^d$ lie in $[0,1]$. A value of $\Lambda_{ij}^d=1$ means that, for example, a 50\% reduction in the production of supplier $i$ can lead to a 50\% reduction in the production of customer $j$. Importantly, for any essential input, the sum of $\Lambda_{ij}^d$ over all suppliers providing that input equals one. Thus, if all suppliers of an essential input cease production, customer firm $j$ must also cease production completely. Non-essential inputs, in contrast, can only cause a complete stop in production if the firm has no essential inputs. We call an entry with $\Lambda_{ij}^d=1$ a high impact link, because the failure of supplier $i$ can, in principle, completely stop production at firm $j$.

An important feature of the ESRI model is that suppliers can be replaced. Small suppliers are generally easier to replace than large suppliers, but replacement becomes more difficult when a shock affects a large fraction of the supplier's industry. To capture this mechanism, we define a supplier-replaceability factor that attenuates the propagation of shocks:
\begin{equation}
    \sigma_i\left(t\right)=\min\left[\frac{s_i}{\sum_j s_j\delta_{p_i p_j}\left(1-x_j^d(t)\right)},1\right]\,.
    \label{eq:Sigma_SI}
\end{equation}
Here, $s_i$ denotes the total value of goods sold by firm $i$, i.e., its out-strength, $s_i=\sum_j W_{ij}$. The variable $x_i^d(t)$ denotes the fraction of production lost by firm $i$ due to input shortages at time $t$. At time $t=0$, before any firm has been affected by a shock, $\sigma_i(0)$ is equal to the market share of firm $i$ within its industry. We therefore refer to $\sigma_i(t)$ as the effective market share of firm $i$ at time $t$. The time dependence of $\sigma_i(t)$ is crucial: even a small firm can become difficult to replace if the rest of its industry is strongly affected by a shock.

We can now compute how an initial shock to firms' production capacities, $\psi$, propagates through the network and causes supply-side constraints for other firms:
\begin{multline}
        x_j^d(t+1) = \max\Biggl[\max_{p_l\in \mathcal{I}^\text{es}_j}\left[\sum_i \Lambda^d_{ij}x_i^d\left(t\right)\sigma_i\left(t\right)\delta_{p_i p_l}\right]\,,\\
    \sum_{p_l\in\mathcal{I}^\text{ne}_j}\sum_i \Lambda^d_{ij}x_i^d(t)\sigma_i(t)\delta_{p_i p_l},\psi_j\Biggr]\,.
    \label{eq:downprop_SI}
\end{multline}
The entry $\psi_j$ of the initial shock vector, $\psi$, denotes the direct shock to the production capacity of firm $j$. A value of $\psi_j=1$ means that firm $j$ completely ceases production as a consequence of the initial shock. The first term captures constraints arising from essential inputs, where the most severe input shortage determines the production loss. The second term captures constraints arising from non-essential inputs, whose effects accumulate additively. The third term ensures that the direct initial shock to firm $j$ is retained throughout the propagation process. Note the nonlinearity introduced by the effective market share, $\sigma_i(t)$.

Shocks can also propagate upstream, from buyers to suppliers. If a buyer reduces its production, it purchases proportionally fewer inputs from its suppliers. This demand-side propagation is described by
\begin{equation}
    x_j^u\left(t+1\right) = \max\left[\sum_i \Lambda^u_{ij}x_i^u\left(t\right),\psi_j\right]\,,
    \label{eq:upprop}
\end{equation}
where the upstream propagation matrix $\Lambda^u_{ij}$ is given by
\begin{equation}
    \Lambda_{ij}^u = \begin{cases}
        \frac{W_{ji}}{\sum_l W_{jl}}\quad\text{ if } W_{ji}\neq0\,,\\
        0\quad\qquad\text{ else.}
    \end{cases}
\end{equation}
Here, $\Lambda^u_{ij}$ measures the fraction of supplier $j$'s total sales that go to buyer $i$. Thus, if buyer $i$ reduces production, supplier $j$ experiences a proportional reduction in demand. Unlike downstream propagation, upstream propagation follows a linear model: it depends only on the share of supplier $j$'s sales accounted for by buyer $i$.

The upstream and downstream propagation equations, Eqs.~\ref{eq:downprop_SI}--\ref{eq:upprop}, are iterated until the change in production capacity of each firm falls below $\epsilon=0.01$ at time $T$. The final output reduction of firm $i$ caused by an initial shock, $\psi$, is then calculated as
\begin{equation}
    x_i\left(\psi\right) = \max\left[x^u_i\left(T\right),x^d_i\left(T\right)\right]\,,
\end{equation}
where $x^u_i(T)$ denotes the relative reduction in output due to reduced demand, i.e., upstream contagion, while $x^d_i(T)$ denotes the relative reduction in output due to limited availability of intermediate inputs, i.e., downstream propagation. The total effect of a given shock, $\psi$, on the network is summarized by the relative network-wide reduction in production and is referred to as ESRI:
\begin{equation}
    \text{ESRI}\left(\psi\right) = \frac{1}{\sum_l s_l}\sum_j s_j x_j\left(\psi\right)\,.
    \label{eq:DefESRI}
\end{equation}
The shock vector, $\psi$, can encode arbitrary firm-level disruption scenarios. A shock vector in which only the $j$'th entry equals one, while all other entries are zero, represents the complete failure of firm $j$. Multiple simultaneous firm failures can be represented by adding the corresponding single-firm shock vectors. The resulting vector therefore encodes a scenario in which all selected firms are shocked simultaneously.

\subsection{Robustness against the chosen thresholds}\label{sec:Sensitivity}
\begin{figure}
    \centering
    \includegraphics[width=\linewidth]{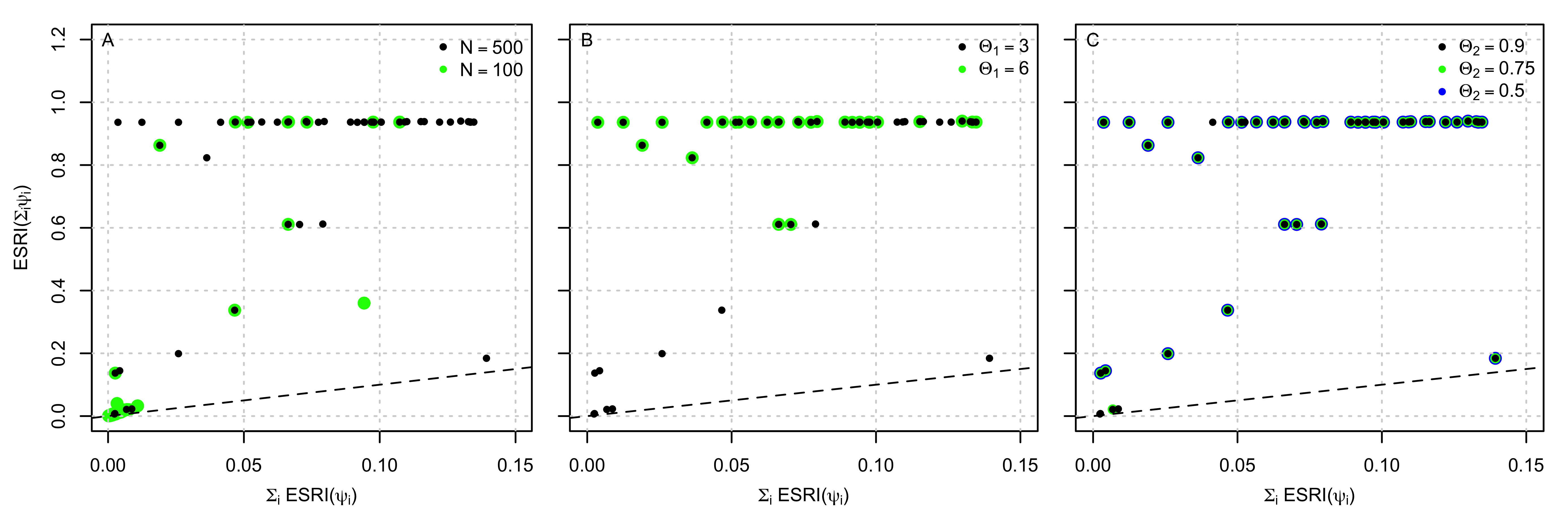}
    \caption{\textbf{Sensitivity analysis of the sampling method} We show scatter plots of the ESRI of the simultaneous failure of identified firm sets against the sum of the ESRIs of the constituent firms. In each plot, we vary one of the parameters of the sampling scheme used to find the sets of firms, and unless otherwise noted, we use 100,000 sets of 500 firms each, with thresholds for identification chosen at $\theta_1=3$ and $\theta_2=0.9$. In panel ,A we use initial set sizes of 100 (green) and 500 firms (black). The larger initial sets are more efficient at identifying the destructive sets. In panel ,B we vary the first threshold from $3$ to $6$. This threshold controls how much larger the impact of a set of 500 firms has to be than the sum of its constituent parts to be considered for further examination. The larger threshold misses dots with a lower amplification factor, but still succeeds at identifying the strongly amplified sets. In panel ,C we examine the impact of the second threshold $\theta_2$ with values of 0.9 (black), 0.75 (green), and 0.5 (blue). Almost all sets are successfully identified regardless of the choice of $\theta_2$. Exceptions occur for very small risk values that are missed by the more stringent threshold and a single pair of firms with a large amplification factor.}
    \label{fig:Sensitivity}
\end{figure}

In Fig.~\ref{fig:Sensitivity} we present the sensitivity analysis we performed for our sampling method. We plot $\text{ESRI}(\sum_i \psi_i)$ against $\sum_i\text{ESRI}(\psi_i)$. Black dots represent small sets identified with a baseline of 100.000 different initial sets of $N=500$ firms each, with detection thresholds set at $\theta_1=3$ and $\theta_2=0.9$. The dashed line represents the diagonal, where the sets are as destructive as the sum of the firms' individual systemic risks.

In Fig.~\ref{fig:Sensitivity}~A, we present the baseline results with black dots and results using 100.000 $N=100$ firm sets in green. Overlapping dots signify sets that are identified in both cases. The extraction algorithm with $N=100$ firms struggles to identify the most destructive sets, i.e., those with $\text{ESRI}\left(\sum_i\psi_i\right)>0.9$, but succeeds in identifying sets that exhibit a small degree of nonlinear amplification with little systemic impact.

In Fig.~\ref{fig:Sensitivity}~B, we present the sets identified with the baseline configuration with black dots and those with $\theta_1=6$ as larger green dots. The overlapping dots in the top left of the plot show that sets with large nonlinear amplification are easily identified independently of the chosen threshold. But choosing a larger threshold can cause the method to miss sets with a large systemic risk, but low amplification factors, as indicated by the black dots in the top right. This is because a large $\theta_1$ can make the threshold, $\theta_1\sum_{i\in \mathcal{S}_k}\text{ESRI}\left(\psi_i\right)$, exceed 1. Since ESRI is bounded above by one, such a threshold would be impossible to reach. A larger threshold for $\theta_1$ further hides sets with low overall systemic impact.

In Fig.~\ref{fig:Sensitivity}~C, we compare the baseline results with those obtained by varying the second threshold $\theta_2$ from 0.9 (black dots) to 0.75 (green dots) and 0.5 (blue dots). Most dots are overlapping, showing that this threshold is largely irrelevant in identifying sets that exhibit nonlinear amplification. This makes sense, since this threshold is used to identify the minimal subsets from the larger 500 firms that lead to nonlinear amplification.

\subsection{Validity check for the Softdrink subnetwork}\label{sec:Validation}
\begin{figure}
    \centering
    \includegraphics[width=0.75\linewidth]{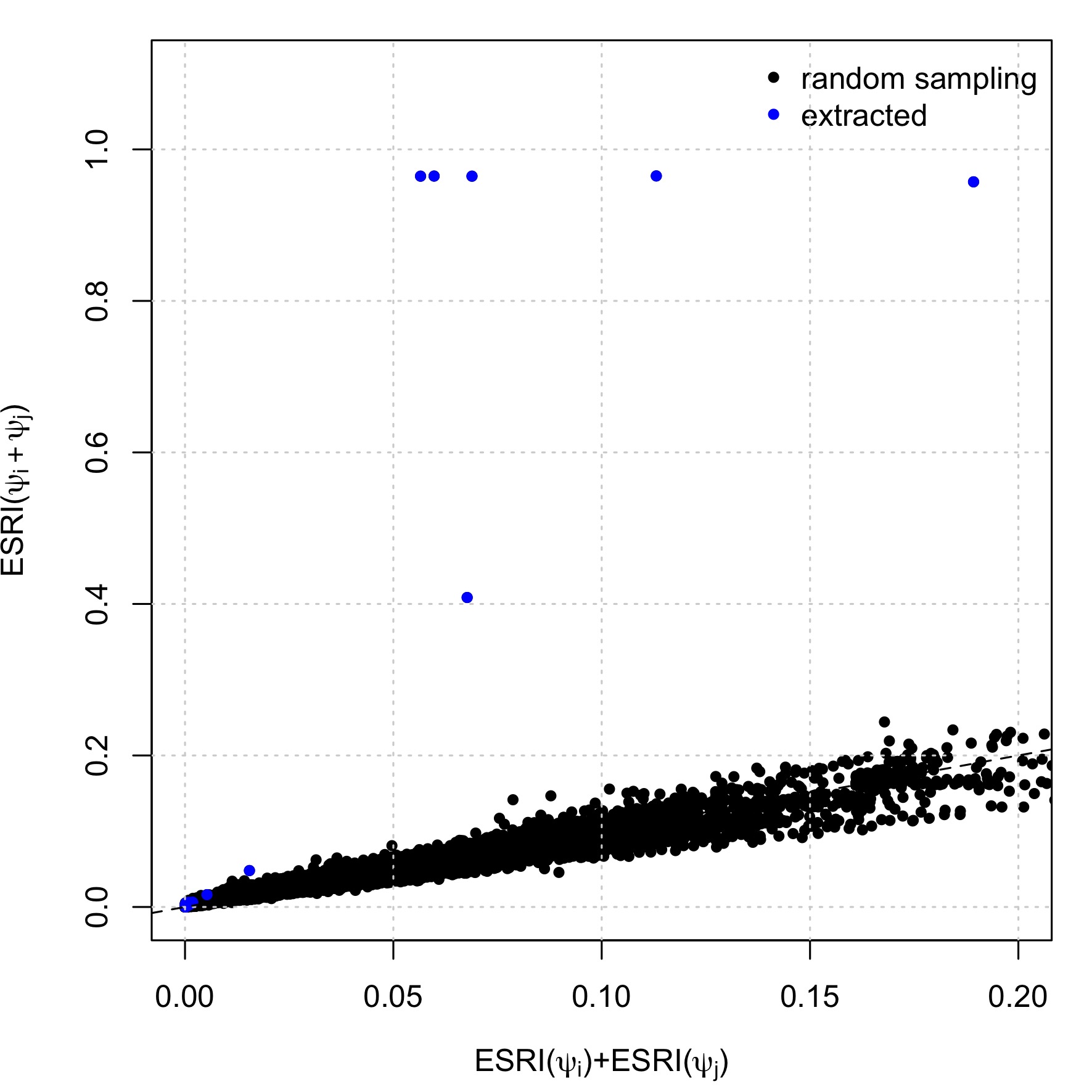}
    \caption{\textbf{ESRI of simultaneous failure against the sum of individual ESRIs for the Softdrink subnetwork} We show the value of the pair failure ESRI against the sum of individual ESRI values. Black dots are the results of an exhaustive search of all possible firm pairs. Blue dots represent node pairs identified with our extraction procedure. Note how, for the large ESRI pairs, our extraction method manages to identify all of them. We use 100,000 sets of $N=5$ firms with $\theta_1=3$ and $\theta_2=0.9$.}
    \label{fig:MethodCheck}
\end{figure}

To test the validity of our extraction method, we use it to examine the softdrink subnetwork, where we can examine every single firm pair exhaustively. We use 100,000 sets of 5 firms with detection thresholds of $\theta_1=3$ and $\theta_2=0.9$ and present the results in Fig.~\ref{fig:MethodCheck}. We plot $\text{ESRI}(\psi_i+\psi_j)$ against $\text{ESRI}(\psi_i)+\text{ESRI}(\psi_j)$ for every possible pair in the softdrink network (black dots) and for the pairs we find with the extraction procedure (blue dots). Our method manages to find all the pairs with a large amplification factor, $\alpha>3$, and a large total systemic impact, $\text{ESRI}(\psi_i+\psi_j)>0.1$.

\subsection{Why those subnetworks?}\label{sec:exportbasket}
\begin{figure}
    \centering
    \includegraphics[width=\linewidth]{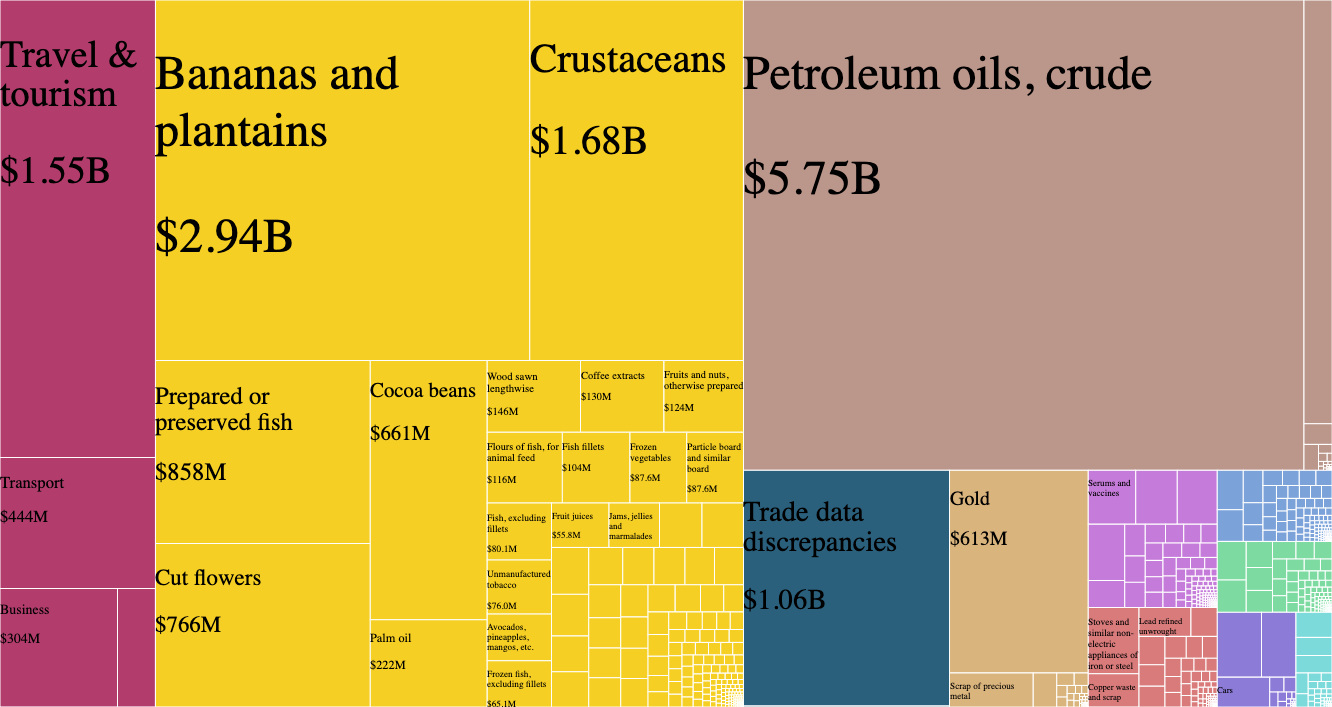}
    \caption{\textbf{Treemap of the export basket of Ecuador for the year 2015 according to the Harvard Growth Lab\cite{Har26}.} The distribution of exports of Ecuador shows that bananas and plantains are the second largest export with a volume of 2.94 billion dollars. The export of bananas is second only to the export of crude petroleum, with a volume of 5.75 billion dollars. The export of crustaceans and prepared or preserved fish is the third largest export group with a total volume of 2.538 billion dollars. The total export volume of Ecuador in the year 2015 was 21 billion \$.}
    \label{fig:HarvardGrowthLab}
\end{figure}

In Fig.~\ref{fig:HarvardGrowthLab}, we show the export basket for Ecuador of the year 2015. The picture is taken from the Harvard Growth Lab \cite{Har26}. Ecuador exported goods with a total value of 21 billion \$ in 2015, of which the largest exports were 5.75 billion dollars of crude petroleum oils (28.02\%). The second largest export group was bananas and plantains with a total volume of 2.94 billion \$ (14.32\%), followed by the exports of crustaceans and prepared or preserved fish with a total volume of 2.538 billion dollars (12.33\%). Together, these categories make up 54.67\% of the total export of Ecuador in the year 2015, so we expect well-developed supply chains amenable to our simple subnetwork extraction method. We focus on the bananas, crustaceans, and prepared or preserved fish supply chains in our extraction procedure, because they make up a large part of Ecuador's export, and firms contributing to those sectors are easy to identify. We do not extract the supply chain network for crude petroleum oils because the oil industry of Ecuador is largely controlled by a single, state-controlled entity.

The crustacean, as well as the banana supply chain networks, are dominated by extraction operations, like fishing or farming, and logistics. To include a supply chain network with goods transforming processes, we extract a supply chain network focused on the softdrink industry. We chose this particular industry because it is easy to identify in the data.
\newpage

\section{Additional Results}
\subsection{ESRI profiles for the 4 SCNs}\label{sec:ESRIs}
\begin{figure}
    \centering
    \includegraphics[width=0.6\linewidth]{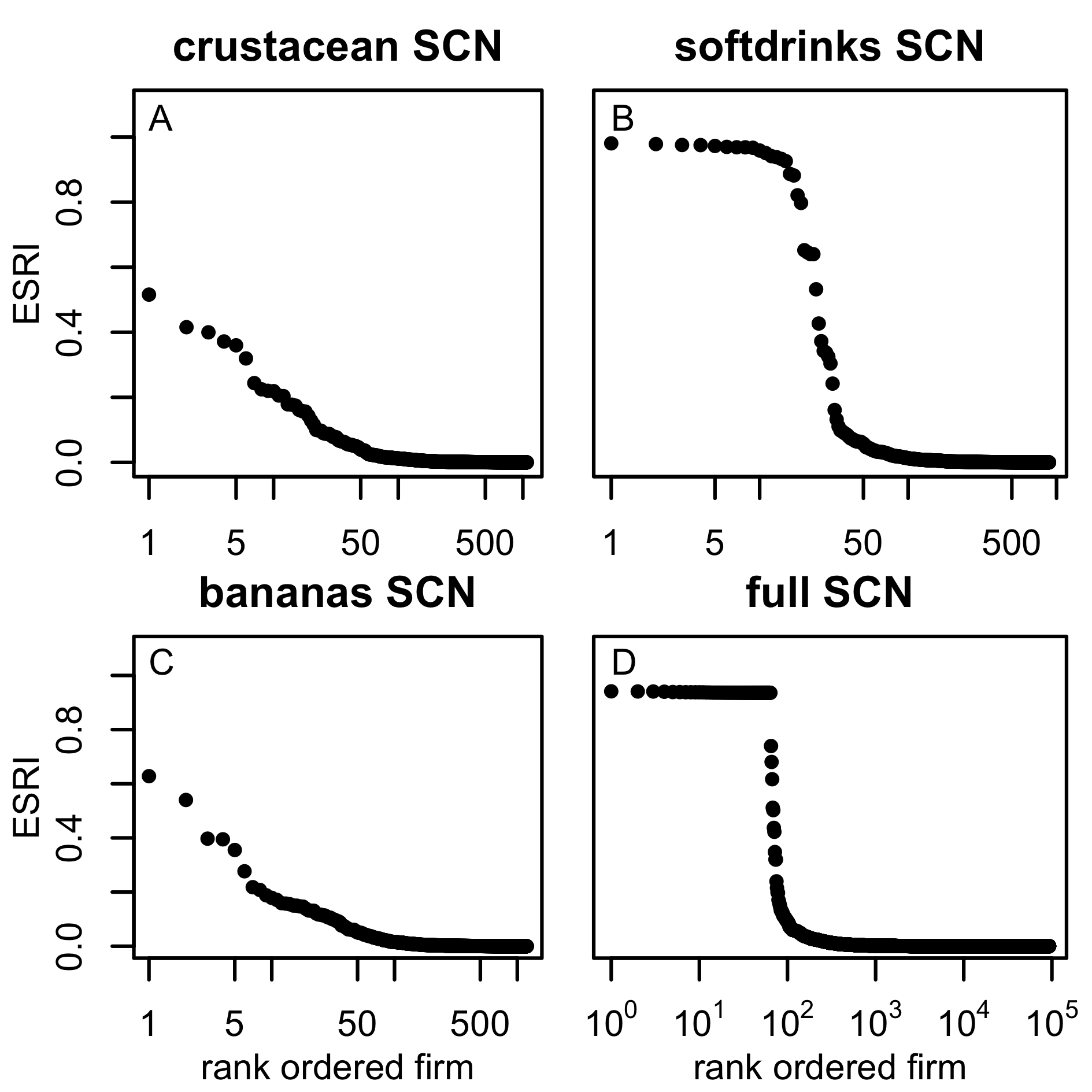}
    \caption{\textbf{ESRI values for single firm failures for the crustacean, softdrink, and banana subnetworks as well as the full supply chain network of Ecuador} The ESRI value for single firm failures, ordered by the rank of each firm, i.e., the most destructive firm is at position 1. Not the log-scale of the x-axis. The left column shows the crustacean supply chain network and the banana supply chain network, with flat profiles that decrease slowly. Here, the most risky firm has an ESRI value of around 0.6. The right column shows the ESRI profiles for the softdrink supply chain network and the full supply chain network of Ecuador, with a pronounced plateau of a few firms (15 and 65, respectively) with an ESRI larger than 0.9, i.e., they reduce the output of the network by 90\% in the case of their failure. The four networks consist of 1075, 1195, 890, and 92,789 firms for the crustacean, bananas, softdrink and the full supply chain network, respectively.}
    \label{fig:ESRI_Profiles}
\end{figure}

To understand the properties of nonlinear amplification, it is helpful to first understand the impact of the failure of single firms on the network. In Fig.~\ref{fig:ESRI_Profiles} we present the single-firm ESRI values for the crustacean, softdrink, and banana subnetworks as well as for the full supply chain network of Ecuador. Each dot presents the ESRI of the failure of a single firm in a rank-ordered manner, i.e., the most destructive firms with the largest ESRI are to the left. Panel A shows the results for the crustacean supply chain network. The most destructive firm has $\text{ESRI}(\psi_i)=0.515$, i.e., it reduces the total output of the network by roughly 50\% should it fail. The rest of the ESRI distribution falls off slowly, dropping to less than 1\% after 114 firms of the 1075 total firms. The banana supply chain network, shown in panel C, shows a similar behavior, the most destructive firm has an ESRI of 0.628, and 125 out of 1195 firms impact more than 1\% of the total network output. The networks in the right column, panels B and D, exhibit a markedly different distribution. Panel B shows the ESRI distribution for the softdrink subnetwork. Here, 15 out of 890 firms affect more than 90\% of the total output of the network should they fail. These firms form a plateau in the distribution of economic systemic risk, and we refer to them as "plateau-firms". In the softdrink supply chain network, a total of 111 firms out of 890 have an ESRI larger than 0.01. The full supply chain network of Ecuador, presented panel D, exhibits a qualitatively similar behavior to the softdrink supply chain network. Here, we find a pronounced systemic risk plateau of 64 firms that affect more than 90\% of the total output of the network in the case of their failure.

\begin{figure}
    \centering
    \includegraphics[width=\linewidth]{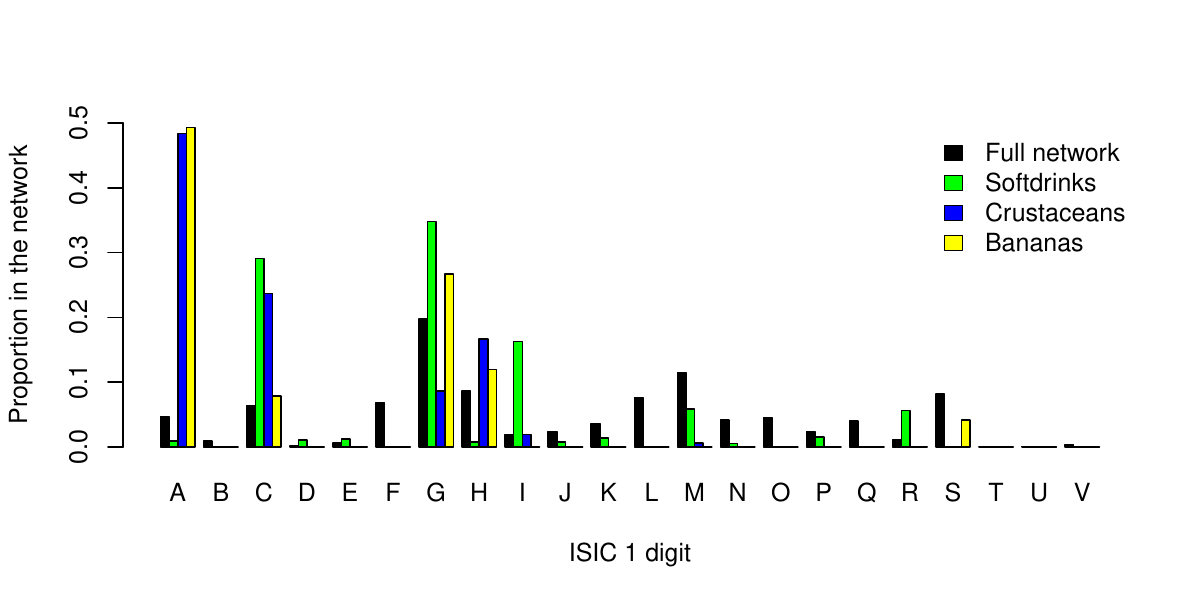}
    \caption{\textbf{Histogram of the industry distribution for the four supply chain networks.} We present a histogram of the industry distribution at the section level according to the ISIC classification scheme. We present the relative number of firms in sections A to V for the full supply chain network of 2015 (black), for the softdrink supply chain network (green), the crustacean supply chain network (blue), and the bananas supply chain network (yellow).}
    \label{fig:Industries}
\end{figure}

Figure~\ref{fig:Industries} shows histograms of the relative number of firms in each industry section for the four supply chain networks. The full supply chain network of Ecuador from 2015 is presented with black bars and serves as a baseline. The softdrink supply chain network is presented as green bars. Note the strong overrepresentation of the manufacturing section "C". The manufacturing sector tends to have more inputs classified as essential, making it sensitive to disruptions. The crustacean and banana supply chain networks are presented in blue and yellow, respectively. Note how half of the firms in those networks are in the aqua- and agriculture section "A". These firms are less prone to disruptions from a lack of intermediate inputs, making these networks less susceptible to systemic risk.

\subsection{Results for the banana subnetwork}\label{sec:Bananas}
\begin{figure}
    \centering
    \includegraphics[width=0.5\linewidth]{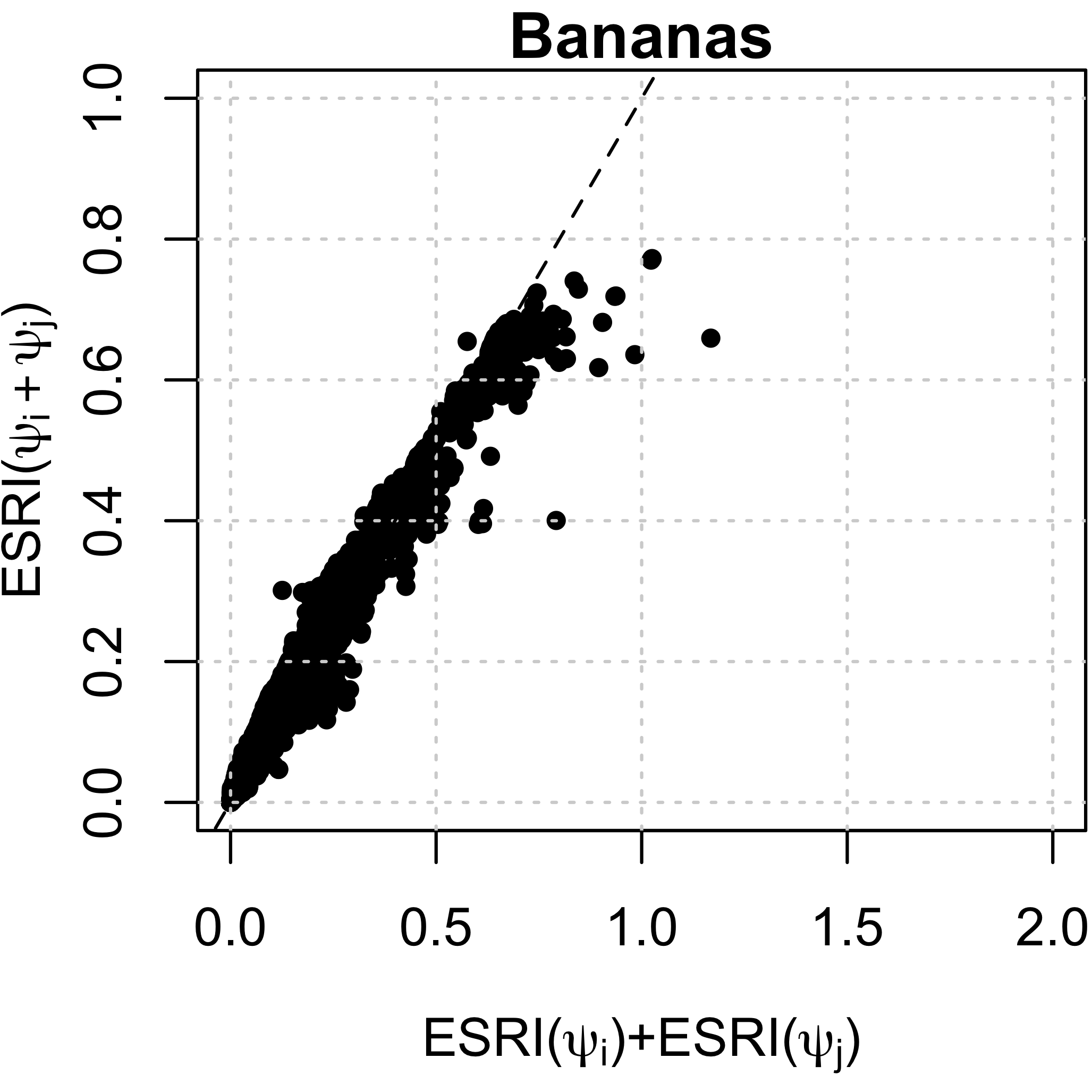}
    \caption{\textbf{Scatter plot of the ESRI for simultaneous failure of firm pairs against the sum of the individual firms' ESRI} The ESRI of simultaneous failure of pairs of firms is plotted against the sum of ESRIs for the firms in the set for the banana subnetwork. Similar to the crustacean subnetwork, no single firm has an ESRI close to 1, and most pairs are close to the identity line, where the sum of their individual ESRIs is comparable to the ESRI of their simultaneous failure. There are some pairs whose simultaneous impact is significantly smaller than the sum of their individual impacts would suggest, and almost none where it is larger.}
    \label{fig:Figure2_Bananas}
\end{figure}

In Fig.~\ref{fig:Figure2_Bananas} we show $\text{ESRI}(\psi_i+\psi_j)$ against $\text{ESRI}(\psi_i)+\text{ESRI}(\psi_j)$ for all pairs of the banana subnetwork. The line represents where the impact of a pairwise simultaneous failure is the same as the sum of the two individual failures. Similar to the crustacean subnetwork, most pairs are on or below the diagonal. This means that nonlinear amplification is very limited in the bananas subnetwork. This lack of nonlinear amplification is to be expected because, like the crustacean subnetwork, the bananas supply chain network does not contain nodes with a large ESRI.

\subsection{Results for the 2010 network of Ecuador}\label{sec:2010network}
\begin{figure}
    \centering
    \includegraphics[width=0.75\linewidth]{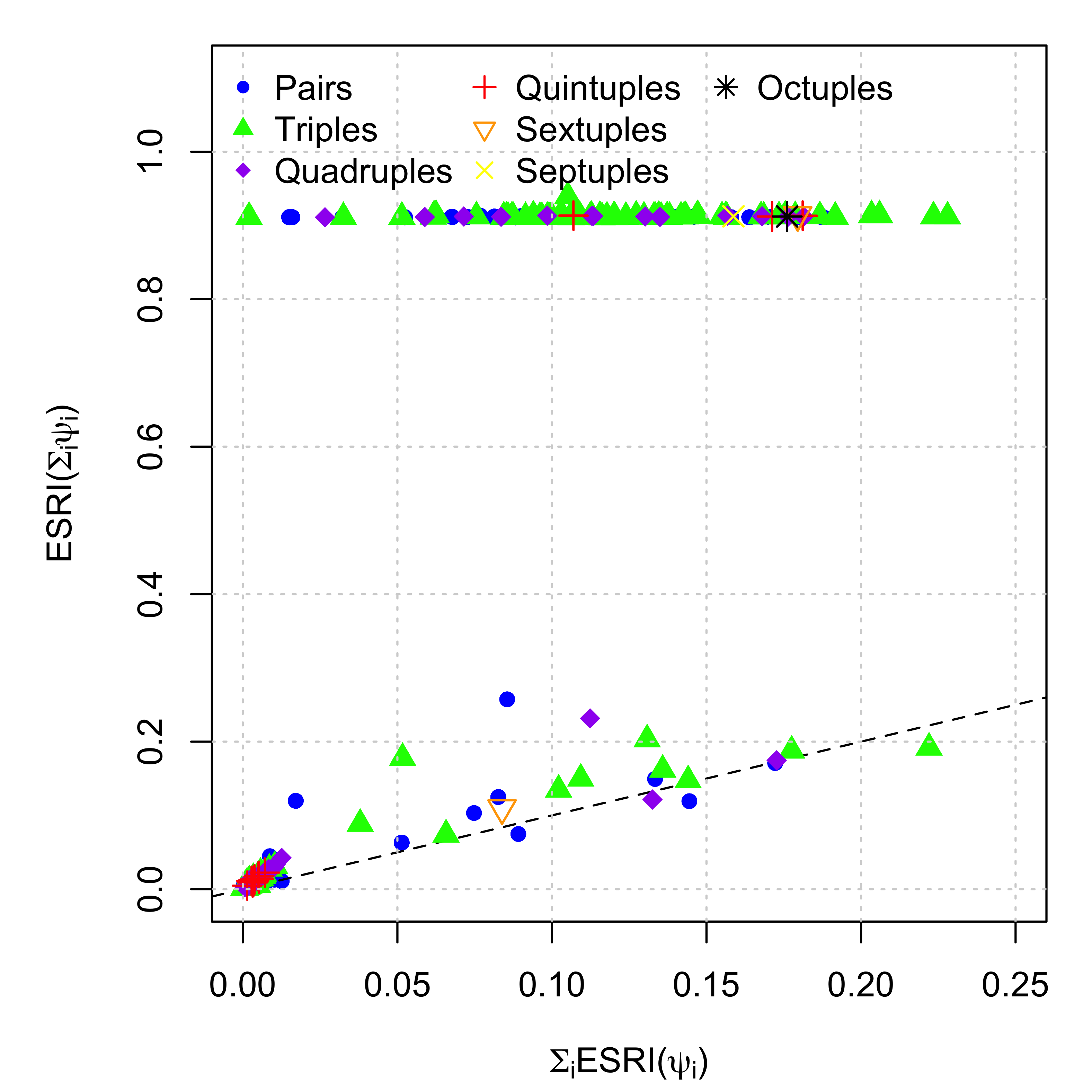}
    \caption{\textbf{Scatter plots of the ESRI for simultaneous firm failures against the sum of individual firms' ESRI} The ESRI of simultaneous failure of small sets of firms is plotted against the sum of ESRIs for the firms in the sets for the full supply chain network of Ecuador in 2010. We also show the identity line as a visual aid. The colored symbols are small sets of firms identified with our extraction procedure. We manage to identify pairs (blue dots), triples (green triangles), quadruples (purple diamonds) and quintuples (red crosses) that are capable of affecting the entire system, even if any individual firms' impact is less than 10\%. We also manage to identify sextuples (orange downturned triangles), septuples (yellow diagonal crosses), and octuples (black stars), whose simultaneous failure is larger than the sum of the failures of their constituents by a factor close to 3, though most of them still have a small impact and are gathered around the origin of the plot. For our sampling method we use 200,000 initial sets of 500 and 100 firms and chose as thresholds $\theta_1=3$ and $\theta_2=0.9$.}
    \label{fig:Figure2_2010}
\end{figure}

We examine the existence of nonlinear amplification over time by repeating the extraction procedure for the dataset from 2010. We use 200,000 initial sets of 500 and 100 firms each, and use the same thresholds as in the main text, $\theta_1=3$ and $\theta_2=0.9$.  We show the results in Fig.~\ref{fig:Figure2_2010}. For sets of up to 8 firms we present $\text{ESRI}(\sum_i \psi_i)$ against $\sum_i \text{ESRI}(\psi_i)$, with the dashed line indicating the diagonal. We identify a total of 138 firm sets, whose simultaneous failure reduces the total output of the network by more than 80\%. This means that the existence of nonlinear amplification is stable over time for the same economies. Interestingly, we identify three pairs that occur in the 2015 and the 2010 networks. In the 2015 network, they exhibit amplification factors $\alpha$ of 14.2, 9.9, and 9.9, and in the 2010 network of 10.6, 13.4, and 8.1. This indicates that the impact on the network of some firm pairs is stable over time.

\subsection{Strong nonlinear amplification is rare}\label{sec:AmpIsRare}
\begin{figure}
    \centering
    \includegraphics[width=0.5\linewidth]{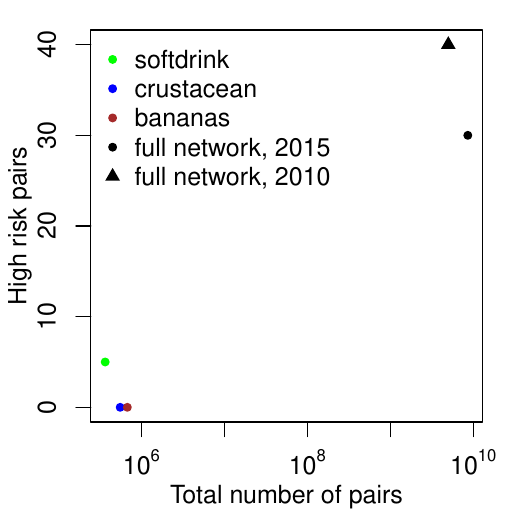}
    \caption{\textbf{Number of firm pairs with $\text{ESRI}(\psi_i+\psi_j)>0.8$ against the total number of pairs of firms with $\text{ESRI}(\psi_i)<0.1$.} For each supply chain network we present the total number of pairs with large systemic risk, $\text{ESRI}(\psi_i+\psi_j)>0.8$ against the total number of pairs in that network in a linear-log plot. The banana (brown) and crustacean supply chain network (blue) contain no firm pair with large systemic risk. The softdrink network contains only five pairs with large systemic risk where the individual firms fulfill $\text{ESRI}(\psi_i)<0.1$. The full supply chain network of Ecuador from the year 2015 (black circle) contains almost $10^{10}$ pairs and we identify 30 pairs with high systemic risk. The full supply chain network of 2010 (black triangle) contains 40 pairs with high systemic risk and $5\times10^9$ total pairs. Note that we only examine consider pairs, where each firm has a low systemic risk $\text{ESRI}(\psi_i)<0.1$ in line with the analysis performed in the main text.}
    \label{fig:riskvstotalpairs}
\end{figure}

In Fig.~\ref{fig:riskvstotalpairs}, we present the number of pairs with high systemic risk, i.e., $\text{ESRI}(\psi_i+\psi_j)>0.8$, against the total number of pairs for five supply chain networks. Note that we restrict pairs to only those where both firms are low risk, i.e., $\text{ESRI}(\psi_i)<0.1$, to be comparable to the results from the main text. The softdrink network is presented as a green circle and contains a total of five high risk pairs and about 350,000 total pairs. The crustacean and banana supply chain networks contain no high risk pairs and a total of about 550,000 and 680,000 pairs, respectively. Neither of these networks exhibits a systemic risk plateau, which means they are robust to firm failure compared to the other three networks, see also SI section~\ref{sec:ESRIs}. The full supply chain networks of Ecuador from 2015 (black circle) and 2010 (black triangle) contain a total of $10^{10}$ and $5\times10^9$ pairs each, but only 30, respectively 40 high risk pairs. While the number of pairs has increased by 4 orders of magnitude, the number of pairs of low risk firms that exhibit a high systemic risk increased by less than one order of magnitude. While high risk pairs can have a disproportionate impact on the supply chain network, they are exceedingly rare.

Because strong nonlinear amplification is very rare, its impact on the average systemic risk of pairwise failure is weak. To confirm this we calculate the average systemic risk of simultaneous pairwise failure, i.e., $\langle\text{ESRI}(\psi_i+\psi_j)\rangle_{ij}$, and compare it with the average of the sum of two single firm systemic risks, i.e., $\langle\text{ESRI}(\psi_i)+\text{ESRI}(\psi_j)\rangle_{ij}$, for the full supply chain network of Ecuador from 2015. Our results are biased towards higher risks, because we explicitly extracted firms with strong nonlinear amplification. We assume that we have identified all pairs that exhibit strong nonlinear amplification with our extraction method. This allows us to enrich our data by using the average systemic risk of pairwise failure, calculated from our uniform random sampling, for the pairs we have not observed. In other words, we assume that our 2,000,000 random samples of firm pairs are representative of all other firm pairs that were not identified with the extraction method. With this, we can calculate
\begin{equation}
    \frac{\langle\text{ESRI}(\psi_i+\psi_j)\rangle_{ij}}{\langle\text{ESRI}(\psi_i)+\text{ESRI}(\psi_j)\rangle_{ij}}=1.0071\,.
\end{equation}
That is, the average systemic risk of two simultaneously failing firms is only 0.71\% higher than the average sum of two random single firm systemic risks. The average impact of nonlinear amplification on average systemic risk is larger for the small subnetworks. While these networks typically exhibit lower amplification than the full networks, the relative frequency of large amplification is much higher. This results in an average increase of systemic risk due to simultaneous pairwise failure of 1.4\%, 1.1\%, and 3.0\% for the crustacean, softdrink and banana supply chain network, respectively.

\subsection{Risk amplification of impactful firm sets}\label{sec:Fig3_SI}
\begin{figure}
\centering
\includegraphics[width=0.5\linewidth]{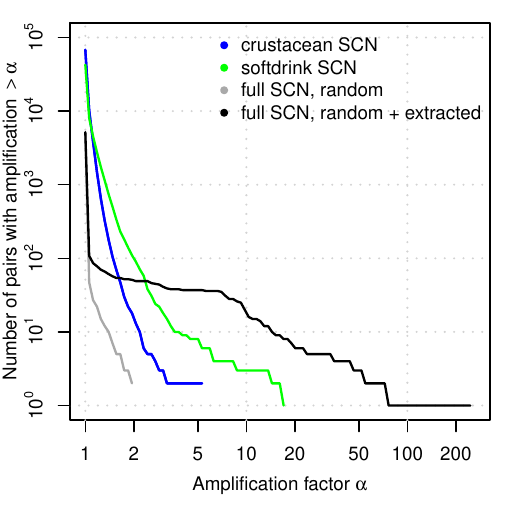}
\caption{\textbf{Survival curves of the amplification factor of simultaneous node failure for pairs with $\text{ESRI}>0.01$.} Shown are the number of firm pairs with an amplification factor larger than a specific value, given that they reduce the network output by at least 1\%. The blue line shows the results for all pairs in the crustacean subnetwork. Of these impactful node pairs, the largest amplification factor is 5.5. The green line represents the amplification factors for impactful pairs in the softdrink subnetwork. The largest amplification of these pairs is 17. The result for the randomly sampled pairs of the full network are shown as a light grey line. Among the highly impactful pairs in that network the largest amplification factor is 2. Our extraction method manages to identify 49 additional pairs with a substantially larger amplification of up to 257. That particular pair has an ESRI sum equal to 0.0036, but their collective ESRI is equal to 0.936.}
\label{fig:Figure3_SI}
\end{figure}

In Fig.~\ref{fig:Figure3_SI} we show the survival function of the amplification factors, $\alpha$, for firm pairs in the different networks, limited to only those pairs with ESRI$\left(\sum_i\psi_i\right)>0.01$. With this, we examine the amplification factors of only those pairs that have an economically significant impact on their respective network. The curves for the crustacean and softdrink subnetworks decrease faster than without that filter. The maximum amplification also decreases to $\alpha=5.5$ and $\alpha=17$ for crustacean and softdrink supply chain networks, respectively. This means most nonlinear amplification occurs for firm sets with a low total impact. This is even more apparent in the full supply chain network, where random sampling finds no firms with $\alpha>2$ and $\text{ESRI}\left(\psi_i+\psi_j\right)>0.01$. This means that the large amplifications observed are limited to low-impact pairs. We manage to identify firm pairs with large amplification factors and large ESRI values with the extraction procedure. Random sampling fails to identify systemically relevant nonlinear amplification, but the extraction method succeeds in identifying these rare but important sets.

\subsection{Network figures of mode examples}\label{sec:ModeExamples}
\begin{figure}
    \centering
    \includegraphics[width=0.5\linewidth]{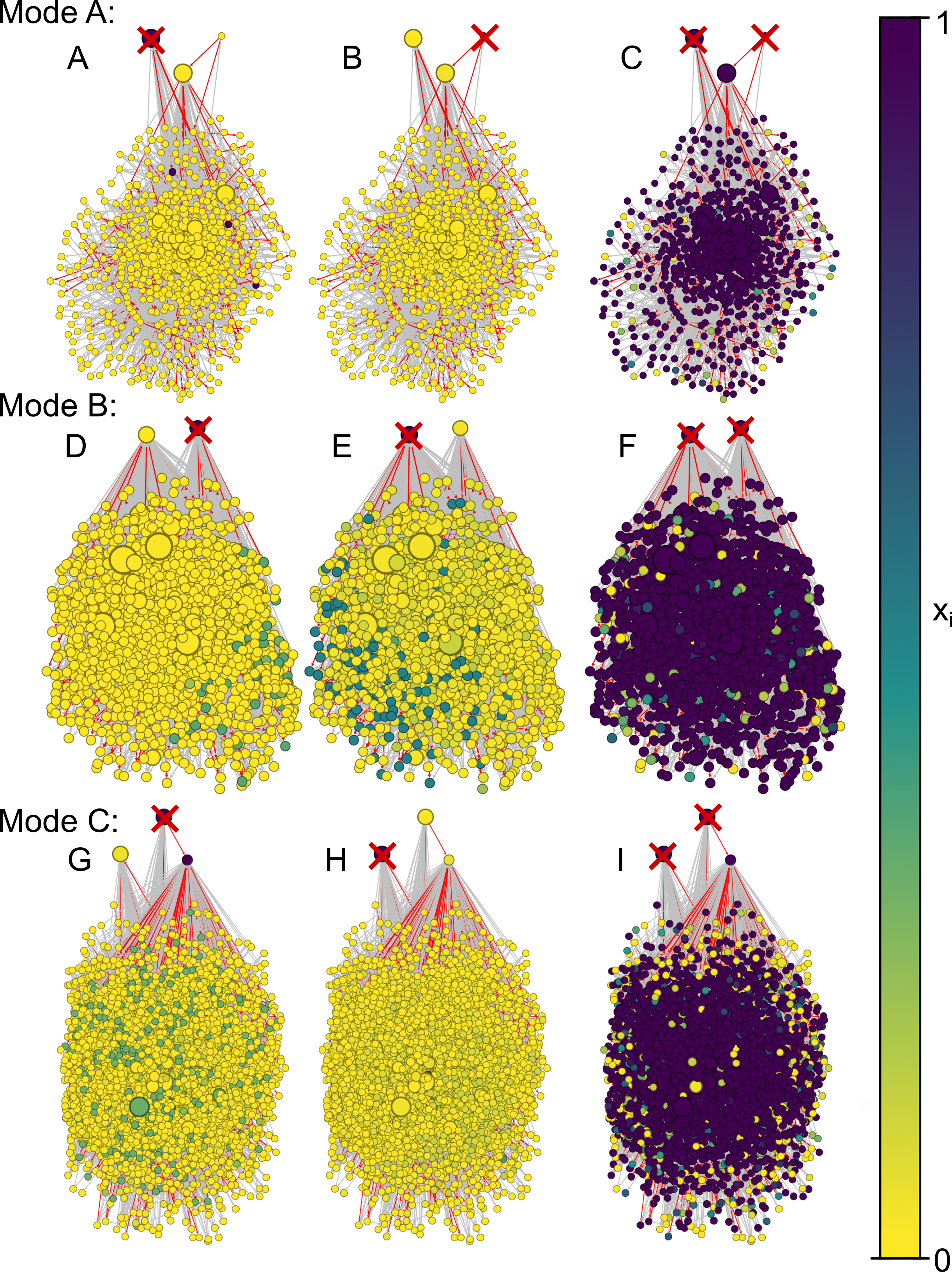}
    \caption{\textbf{Network visualizations of example pairs for the three identified modes.} Panels A -- C show an example pair of the first mode as well as their tier one and two customers. The additional node between the pair and the rest of the network is the "plateau-firm" that is supplied by the smaller firm. Color denotes the size of the production shock, $x_i$, for each firm, and size is proportional to market share. Red links denote high impact links with $\Lambda_{ij}=1$. Panel A shows the result of a shock (red cross) to a firm with a large market share, but little systemic impact. Panel B shows the result of a shock to a firm with a small initial market share that has a high impact link to a "plateau-firm." In Panel C, we present the result of a simultaneous shock to both firms. Panels D -- F show an example pair for the second mode and their direct customers. Panels D and E show the result of shocking only one firm of the pair, and Panel F shows the results of a simultaneous shock to both firms. Panels G -- I present an example pair for the third mode and its first and second tier customers. In panel G, we show the result of a shock to the first firm in the pair. The shock leads to a large disruption in a smaller firm in the same industry as the second firm. This smaller firm has many high impact links to the rest of the network, including a "plateau-firm." In panel H, we show the result of a shock to only the second firm of the pair. Panel I presents the results of a simultaneous shock to both firms in the pair. Note that the third firm is not shocked directly, but is disrupted indirectly through the first firm of the pair.}
    \label{fig:ModeExamples}
\end{figure}

Figure~\ref{fig:ModeExamples} illustrates examples for the three structural modes we identified in Fig.~\ref{fig:Motiveschemas} of the main text. Node size is proportional to a firm's market share and its color represents the final output reduction after an initial shock, $x_i$. Red links denote high impact links, those with $\Lambda_{ij}=1$. Panels A--C present the resulting network state after shocks to a firm pair in the first structural mode. We show the pair itself at the top, the relevant "plateau-firm" in the middle and the first and second tier customers of the pair at the bottom. This pair has the largest observed amplification factor, $\alpha=257$. The two firms have initial market shares of $\sigma_i(0)=94.3\%$ and $\sigma_j(0)=2.4\%$ in sector C3311, “Repair of fabricated metal products.” The failure of only the first firm, shown in Panel A, leads to the complete cessation of production for a few customers, but does not propagate any further. The failure of only the small firm (Panel B) leads to little further effect on the network. This is because its small market share makes it easy to replace. In Panel C, both firms fail simultaneously, and the smaller firm’s effective market share rises to $\sigma_j(t_1)=73.7\%$. This makes it much harder to replace. As a result, the "plateau-firm" experiences a more stringent reduction in the availability of the input supplied by the smaller firm and fails. This leads to a much larger cascade than if the small supplier failed on its own.

The second mode is exemplified in Fig.\ref{fig:ModeExamples}~D--F. The two firms are classified as part of the "Risk and damage evaluation" sector, K6621. They have initial market shares of $\sigma_i(0)=26.3\%$ and $\sigma_j(0)=33.6\%$, and together they are connected by 175 high impact links to firms that account for almost 40\% of the out-strength of the "Activities of insurance agents and brokers" sector, K6622. The two firms are shown as nodes at the top of the panels and their direct customers at the bottom. The failure of only one of these firms, shown in panels D--E, disrupts their customers, but these disruptions do not spread further through the network. The simultaneous failure of both firms, shown in panel F, triggers a cascade that spreads through large parts of the insurance sector and then propagates further through tee rest of the network. Another example of this mode is the scenario shown in Fig.~\ref{fig:Figure1}~D--F of the main text.

An example for the third mode is shown in panels G--I of Fig.~\ref{fig:ModeExamples}. The pair consists of two firms in different industries. The first firm belongs to the "Manufacture of steam generators, except central heating hot water boilers" sector, C2513. Its failure, shown in panel G, reduces the total output of the "Manufacture of soft drinks; production of mineral waters and other bottled waters" sector, C1104, by 27\% and completely stops production at one high impact supplier of a "plateau-firm," the third node on the bottom right of the pair. The second initially shocked firm belongs to sector C1104 and has an initial market share of 60\%. The result of its individual failure is shown in panel H. It causes only a small cascade through the supply chain network. When both firms fail simultaneously, the shared disruption of sector C1104 becomes large enough that supplier replacement breaks down. As a result, the disruption of the indirectly shocked firm can not be attenuated, and large parts of the network, including a "plateau-firm," fail.

\subsection{DebtRank shows no nonlinear amplification}\label{sec:DebtRank}
\begin{figure}
    \centering
    \includegraphics[width=0.5\linewidth]{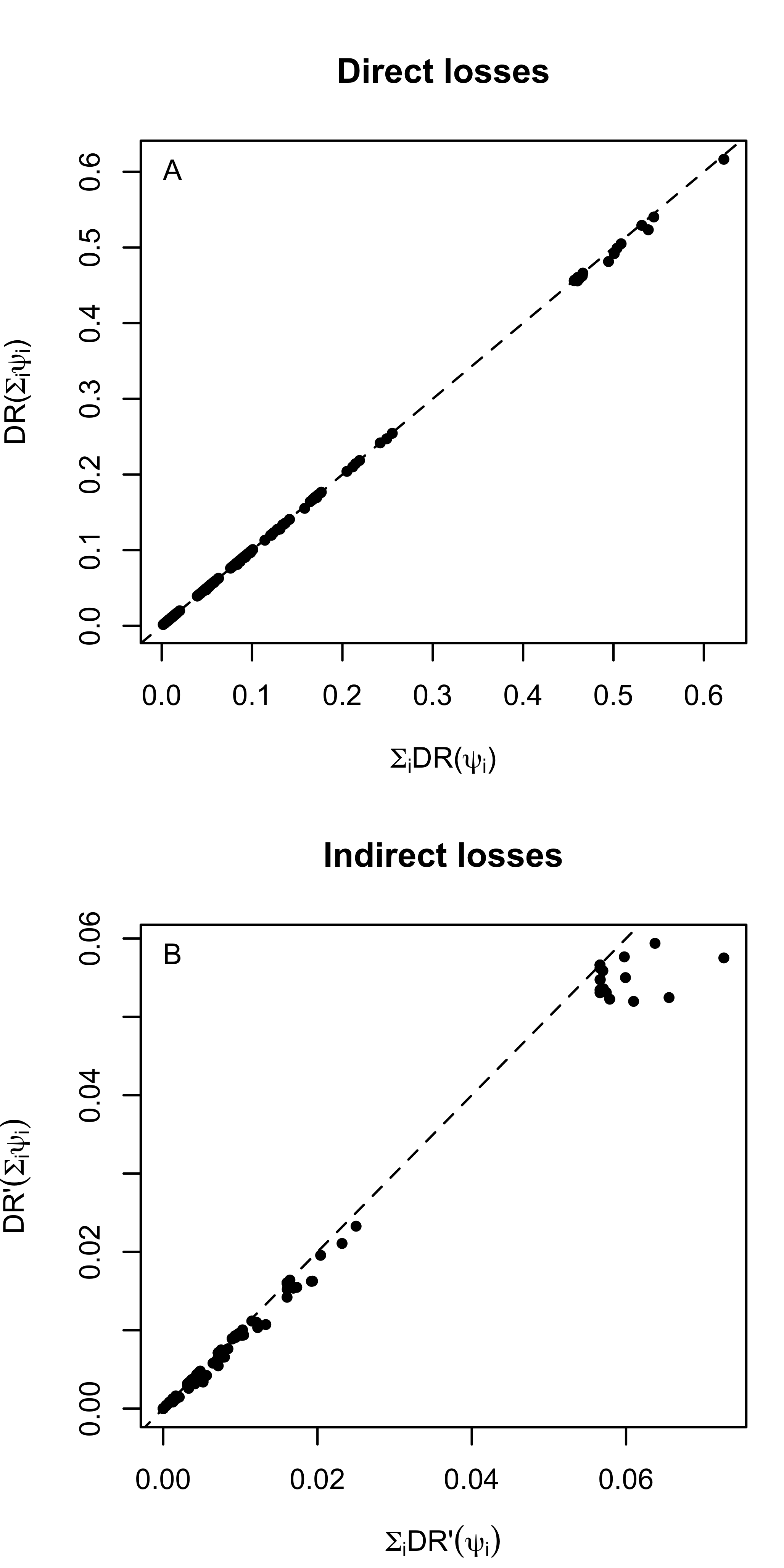}
    \caption{\textbf{Scatter plot of the DebtRank losses from simultaneous pairwise failure plotted against the sum of individual DebtRank.} With a statistical reconstruction of an interbank network with 19 banks, see \cite{Fia25}, we simulate the DebtRank algorithm for each individual bank and for all pairs of banks. In panel A we show the total equity losses in the financial system due to the simultaneous pairwise failure of two banks against the sum of the banks' individual DebtRanks. Note that all dots lie below the dashed diagonal, indicating that the systemic impact of simultaneous failure is always lower than the sum of individual impacts. In panel B we present the same simulations, but without the losses stemming from the equity of the initially failing banks, i.e., we plot only the indirect losses caused by the DebtRank contagion mechanism. This makes the sublinear behavior of DebtRank, as a measure of financial systemic risk, even more apparent.}
    \label{fig:DebtRank}
\end{figure}

One of the most studied measures for financial systemic risk is DebtRank \cite{Bat12}. The central idea of DebtRank is that losses of a bank can affect other banks through a devaluation mechanism of interbank loans. That is, if the debtor, $i$, of a bank, $j$, loses equity, that affects the book-value of that loan. This valuation adjustment directly affects the equity of $j$, thereby continuing the propagation of the original shock. In the simplest case, this adjustment is proportional to the lost equity:
\begin{align}
x_i(t)&=\min\left[1,\; x_i(t-1)+\sum_{j~:~s_j(t-1)=D} W_{ji}x_j(t-1)\right]\,,\\
s_i(t)&=
\begin{cases}
D & \text{if } x_i(t)>0;\, s_i(t-1)\neq I\,,\\
I & \text{if } s_i(t-1)=D\,,\\
s_i(t-1) & \text{otherwise.}
\end{cases}
\end{align}
Here, $x_i(t)$ is the relative amount of equity lost of bank $i$, note the similarity to the reduction in production for the ESRI model. $W_{ij}$ is a matrix that describes the interbank loan exposures for each bank. A value of $W_{ij}=1$ means that bank $j$ would lose all of its equity if the loan it extended to bank $i$ has to be written off completely. Further, $s_i$ is a state variable that prevents each bank from propagating the shock more than once. At time $t=T$, the shock has propagated through the entire interbank network, and the final loss can be calculated as
\begin{equation}
    \text{DR}(\psi) = \frac{\sum_ih_i(T)e_i}{\sum_i e_i}\,,
\end{equation}
where $e_i$ is the equity of bank $i$ and $\psi_i$ is the initial shock.

In Fig.~\ref{fig:DebtRank}~A we present the $\text{DR}(\psi_i+\psi_j)$ against $\text{DR}(\psi_i)+\text{DR}(\psi_j)$ for an interbank network of 19 banks, taken from \cite{Fia25}. The dashed line represents the curve where the two pairwise failures cause the same total loss as the sum of individual failures. All dots are either on top of the line or slightly below it, indicating that DebtRank is purely sublinear for combined failures. In panel B, we show only the indirect losses caused by the initial failure of two banks. That is, we remove the equity of the originally failing banks from the final losses in the interbank network, $\text{DR}^\prime(\psi_i)=\text{DR}(\psi_i)-\frac{e_i}{\sum_i e_i}$. It is readily apparent that the indirect losses are sublinear for simultaneously failing banks. The nonlinearities in DebtRank are the minimum function, which limits the losses of each bank to be at most equal to their equity, and the state variable that limits the shock spreading. Neither of these mechanisms can increase losses in a superlinear fashion; consequently, the observed sublinearity is expected.

\bibliographystyle{unsrtnat}
\bibliography{Bibliography}

\end{document}